\documentclass[useAMS,usenatbib]{mn2e}
\usepackage{graphicx}
\title[VLT-FORS2 imaging and spectroscopic observations of SDSS type 2 quasars]{Interactions, star formation and extended nebulae in SDSS type 2 quasars at 0.3$\leq z \leq$0.6\thanks{Based on observations carried out at the
European Southern Observatory (Paranal, Chile) with FORS2 on VLT-UT1 (programmes 081.B-0129,
 083.B-0381, 087.B-0034).}}
\author[Villar-Mart\'\i n et al.]{M. Villar-Mart\'\i n$^{1}$, C. Tadhunter$^{2}$, A. Humphrey$^{3}$, R. Fraga Encina$^{4}$ \\
\newauthor R. Gonz\'alez Delgado$^{1}$, M. P\'erez Torres$^{1}$, A. Mart\'\i nez-Sansigre$^5$\\
$^{1}$Instituto de Astrof\'\i sica de Andaluc\'\i a (CSIC), Glorieta de la Astronom\'\i a s/n, 18008 Granada, Spain. montse@iaa.es \\
$^{2}$Dept. of Physics and Astronomy, University of Sheffield, Sheffield S3 7RH, UK\\
$^{3}$Instituto Nacional de Astrof\'\i sica, Optica y Electr\'onica (INAOE), Aptdo. Postal 51 y 216, 72000 Puebla, Mexico\\
$^4$Instituto de F\'\i sica de Cantabria (CSIC-UC), Avda.d e los Castros, s/n, E-39005,  Santander, Spain \\
$^{5}$Institute of Cosmology and Gravitation, Univ. of Portsmouth, Dennis Sciama Building, Burnaby Road, Portsmouth, PO1 3FX, UK}

\begin{document}

\date{} 

\pagerange{\pageref{firstpage}--\pageref{lastpage}} \pubyear{2002}

\maketitle

\begin{abstract}
We present  long-slit spectroscopy and  imaging data obtained with FORS2 on the Very Large Telescope  of 13 optically selected  type 2 quasars at $z\sim$0.3-0.6
from the original sample of Zakamska et al. (2003). The sample is likely to be affected by different selection biases. We investigate the evidence for: a) mergers/interactions b) star formation activity in the neighborhood of the quasars and
c) extended emission line regions and their nature.  Evidence
for mergers/interactions is found in 5/13 objects. This
is a lower limit for our sample, given the shallowness of most of our continuum images. Although AGN photoionization cannot be totally discarded, line ratios consistent with stellar photoionization are found in general in 
companion galaxies/knots/nuclei  near these same objects. On the contrary, the gas in the neighborhood of the quasar nucleus shows line ratios inconsistent with HII galaxies and typical of AGN photoionized nebulae. A natural scenario to explain the observations is that  star formation is ongoing in companion galaxies/knots/nuclei, possibly triggered by the interactions.
These systems are, therefore, composite
in their emission line properties showing a combination of AGN and
star formation features. 

 Extended emission line regions (EELRs)
have been found in 7/13  objects, although this fraction might be higher if a complete  spatial coverage around the quasars was performed. The sizes vary between few and up to 64 kpc. In general, the EELRs apparently consist of an
extended nebula associated with the quasar. In at least one case the EELR is associated with ionized tidal features. 

\end{abstract}


\label{firstpage}

\section{Introduction}
According to the standard unification model of active galaxies (AGNs)
certain classes of type 1 and type 2 AGNs are the same entities, but have different orientations relative to the 
observer's line of sight (e.g. Seyfert 1s and 2s; Antonucci \citeyear{ant93}). An obscuring
structure blocks the view to the inner nuclear region
 for some orientations.
If this model is valid for the most luminous AGNs (quasars), there
must exist a high-luminosity family of type 2 quasars. 
The existence of this 
 object class was predicted a long time ago, but it has been only in the last few years that type 2 quasars  have been discovered
in large quantities (e.g. Zakamska et al. \citeyear{zak03}, Mart\'\i nez-Sansigre et al. \citeyear{san05}). 
 In particular, Zakamska et al.
 identified in 2003 $\sim$300 objects in the redshift  range
0.3$\le z \le$0.8 in the Sloan Digital Sky Survey (SDSS)  with the high ionization
narrow emission line spectra characteristic of type 2 AGNs and  narrow
line luminosities typical of type 1 quasars. 
 Their far-infrared luminosities place
them among the most luminous quasars at similar $z$. 
They  show a wide range of X-ray luminosities and obscuring column densities
(Ptak  et al. \citeyear{ptak06}; Zakamska  et al. \citeyear{zak04}). The host galaxies are ellipticals with irregular morphologies  (Zakamska et al. \citeyear{zak06}), and
the nuclear optical emission is highly polarized in some cases (Zakamska et al. \citeyear{zak05}). The detection rate
 at 1.4GHz   in the FIRST survey is $\sim$49\% (Zakamska et al. \citeyear{zak04}).   \cite{lal10} found a detection rate  $\sim$59\% at 8.4 GHz
  using nearly one
order of magnitude deeper VLA images.
 15\%$\pm$5\% qualify as radio loud.

 We are undertaking an imaging and spectroscopic observational programme  using  FORS2 on
VLT  to perform detailed optical imaging and spectroscopic studies of SDSS type 2 quasars and address several critical
 aspects related to the nuclear activity 
 and the evolution of massive elliptical galaxies.

1) {\it The triggering mechanism of the nuclear activity}. Both models  and observations of
powerful type 1 quasars  and radio galaxies support the idea that major mergers and interactions
are an efficient mechanism to trigger the nuclear activity in the most powerful AGN (e.g. Ramos Almeida et al. \citeyear{ram11}) although some authors challenge this interpretation (e.g. Cisternas et al. \citeyear{cis11}). We are  addressing this issue by searching for signatures
of mergers/interactions in the environment of SDSS type 2 quasars.

2) {\it The existence of associated extended emission
line regions (EELRs)}.     Previous works  show that 
luminous  EELRs (L[OIII]$_{EELR}\ge$5$\times$10$^{41}$ erg s$^{-1}$) around type 1 quasars  at  $z\le$0.5 exist preferentially associated with
 steep-spectrum radio-loud quasars with luminous  nuclear emission lines (L[OIII]$_{nuc}$$\ge$6.5$\times$10$^{42}$ erg s$^{-1}$) and low broad line region (BLR hereafter) 
metallicities ($Z\le$0.6$Z_{\odot}$, although it must be kept in mind that metallicity estimations are highly uncertain). The EELRs have low metallicities  and they also seem specially prevalent in quasars showing signs of strong galaxy interactions (e.g. Fu \& Stockton \citeyear{fu09}, \citeyear{fu07}). 
The more frequent 
occurrence of luminous EELRs in  
 radio loud objects  suggests that their origin  might be connected with the radio activity (ejection?)
or, alternatively, that the radio activity is only triggered in rich gaseous environments.
Since  very little is known about the  existence of EELRs in type 2 quasars,
we are addressing this issue in our observational programme by searching for such nebulae and characterizing
their general properties. The SDSS type 2 quasar sample is good for testing the potential importance of 
nuclear emission line luminosity on the existence of EELR, since they are mostly radio quiet and have large nuclear line luminosities.
\cite{green11} have recently found   that ionized gas is spatial scales $\ga$10 kpc is rare in SDSS type 2 quasars at surface brightness levels $\sim$10$^{-16}$ erg s$^{-1}$ cm$^{-2}$ arcsec$^{-2}$. Our data reach surface brightness levels $\ga$20 times deeper.

3)  {\it Quasar feedback processes}. Quasar-induced
outflows are recognized as an important feedback mechanism on the formation 
of massive elliptical galaxies. 
 Studies of powerful radio loud AGNs suggest that
 radio-jet induced outflows can have enormous energies sufficient to eject a large fraction of the gaseous galaxy content (Nesvadba et al.  \citeyear{nes06},  
Humphrey et al. \citeyear{hum06}). However, the majority  of quasars are radio quiet (e.g. Jiang et al. \citeyear{jiang07}). How  AGN  feedback
works in these objects is still an open question. 

Type 2 quasars are unique laboratories to investigate these issues.  The  occultation
of the active  nucleus by the obscuring torus
allows a detailed study of many properties  of the surrounding medium, which
is difficult in type 1 quasars due to the dominant contribution of the quasar PSF. On the other hand, the radio  activity in radio loud quasars and radio galaxies  often imprints  important distortions on the 
 observed optical properties via jet-gas interaction. This complicates the  characterization of  the intrinsic properties of the host galaxy and environment.
Type 2 quasars can help to identify and understand those phenomena that are not related to the radio activity.

 In \cite{vil10} we published the first 
results of our VLT programme for SDSS J0123+00  at $z$=0.4.  We discovered
that this  quasar, which
is a member of an interacting system, is physically connected to a companion galaxy by a $\ge$100 kpc filament which is  possibly forming stars actively.  The origin
of the EELR in this particular case  is a galaxy interaction and our results showed that luminous EELR can also be associated with low radio luminosity quasars, contrary to expectations based on type 1 quasar studies.

 We present in this paper the results based on VLT spectroscopic and imaging data of 
13 more  SDSS type 2  quasars regarding  issues 1) and 2) mentioned above. A detailed study of feedback mechanisms in this sample will
be presented in Villar-Mart\'\i n et al. (2011, in prep). 

We assume
$\Omega_{\Lambda}$=0.7, $\Omega_M$=0.3, H$_0$=71 km s$^{-1}$ Mpc$^{-1}$.

\section{Sample selection, observations and data reduction}

The data  were obtained with the Focal Reducer and Low Dispersion Spectrograph (FORS2) for the Very Large Telescope (VLT) installed on UT1
(Appenzeller et al. \citeyear{appen98}). 
The observations were performed in two different runs: 8 and 9th Sept 2008  (4 objects) and 17 and 18 April 2009 (9 objects). The broad band image of SDSS J1307-02 presented in this paper was obtained on 28 April 2011.

 Three criteria were common to all 13 objects studied in this paper: 

1) they are  type 2 quasars extracted from  Zakamska et al. (\citeyear{zak03}) SDSS sample. They are radio quiet in general ($P_{5~GHz}<$10$^{31}$ erg s$^{-1}$ Hz$^{-1}$ sr$^{-1}$ according to the criteria by Miller, 
Peacock \& Mead \citeyear{mill90}. We assume $P_{\nu} \propto \nu^{-\alpha}$, with $\alpha$=0.7). The only exception is SDSS 1228+00,
which has $P_{5~GHz}=$2.4$\times$10$^{31}$ erg s$^{-1}$ Hz$^{-1}$ sr$^{-1}$, in the transition range  between radio quiet and radio loud objects ($P_{5~GHz}>$10$^{32}$ erg s$^{-1}$ Hz$^{-1}$ sr$^{-1}$).

 2) The SDSS optical spectra show  emission lines with large equivalent widths. In this way we
avoid in most cases uncertainties due to underlying stellar absorption and ensure a high S/N ratio for the kinematic analysis. Objects 
with low line luminosities were {\it not} excluded.  Because of their selection criteria, type 2 quasars  in Zakamska et al.  (\citeyear{zak03})
original sample have  L[OIII] luminosities in the range
  $\sim$[10$^8$-10$^{10}$] L$_{\sun}$. The  L[OIII]range in our sample is [1.2$\times$10$^8$-4.1$\times$10$^{9}$] L$_{\sun}$ as measured from the SDSS spectra.

3) All objects have  redshift $z\sim$0.3-0.6 
such that there was an adequate narrow or intermediate band FORS2  filter containing one of the strongest emission lines in the optical spectrum ([OII]$\lambda$3727 or [OIII]$\lambda$5007).

Additional criteria were applied to 10/13 objects in the sample:  they had prior evidence for mergers/interactions (1 object) and/or  additional interesting data already published (HST, Spitzer, etc) (6 objects) and/or evidence for  perturbed  nuclear kinematics (FWHM of the forbidden lines $>$800 km s$^{-1}$) based on optical
SDSS emission line spectrum (6 objects). We indicate in the last column of Table 1  which of  these additional criteria were applied for each object. ``IS'', ``OD'' and ``PK''  mean 
"Interacting System", ``Other Data'' and ``Perturbed Kinematics'' respectively. A ``--'' in that column means that no additional criteria were applied, apart from  the three mentioned above,
which were  common to all objects.   The selection criteria applied  by other authors
in their  programmes are unknown. Therefore,
our sample is likely to be affected by different biases and the results obtained with our work cannot be extrapolated to
the general population of optically selected type 2 quasars.

The observing programme consisted of obtaining for each object a continuum+emission line image followed by long slit spectroscopy. Besides the interest of their scientific content, 
the goal of obtaining these images was to identify extended ionized nebulae and/or other interesting gaseous features such as tidal tails, compact knots, etc which  
helped to decide how to place the spectroscopic slit.

Unfortunately, due to technical problems at the observatory   such deep continuum+emission line images could only be obtained during the 2008 run (i.e. for 4 objects) and only shallow continuum images are available for the 2009 run (9 objects). As a consequence,  we did not have prior information about the possible existence of extended diffuse emission line  structures and their morphology for most objects observed during this run. The continuum images were not deep enough in general either to look for interesting features such as tidal tails, bridges, etc.  So, the spectroscopic slit had to be placed blindly or through continuum sources near the quasar in the image, with
the goal of checking their $z$.

All spectra  were obtained  with the 600RI+19 grism and the   GG435+81 order sorting filter. The useful spectral range was $\sim$5030-8250 \AA\ in the
2008 run and $\sim$5300-8600 \AA\ in the 2009 run, so that in all cases at least the H$\beta$ and  [OIII]$\lambda\lambda$4959,5007 lines were within the observed spectral range.  The imaging and spectroscopic data reduction process is described in Villar-Mart\'\i n et al. (\citeyear{vil10}). The pixel scales are 0.25$\arcsec$ pixel$^{-1}$ and 0.83
\AA\ pix$^{-1}$ in the spatial and spectral directions respectively. The spectral resolution, as measured from the sky emission lines, was  
  FWHM=7.2$\pm$0.2 \AA\ and 5.4$\pm$0.2 \AA\  for the 2008 and 2009 runs respectively. The slit width was 1.3$\arcsec$ in 2008 and
 1.0$\arcsec$ in 2009.

The log of the observations is shown in Table 1.   The seeing size was measured for each object using several stars in the image.  Comparison of several spectrophotometric standard stars taken with a 5$\arcsec$ slit during the run gave a  flux calibration
accuracy of 5\% over the entire spectral range. 

Using the spectra of several standard stars, geometric distortion was found to be $<$2 pixels (0.5$\arcsec$) across the entire spectral range in all cases, with different values for different 
stars. In order to correct for this effect for each quasar, observations of a bright star with a similar telescope position  would be required. Since these are  not available and 
given the small magnitude of the distortion, we decided not  to apply any correction (the quasar continuum cannot be used for this purpose, because the continuum spatial centroid at different wavelengths can
change for various reasons such as reddening). We were careful when extracting the spectra from different apertures for a given object to make sure that such distortion did not have any negative impact on the analysis.

To perform the spectroscopic analysis, the line profiles were fitted with 
Gaussian functions (one or more, depending on the quality of the fit).
The FWHM values were corrected for instrumental broadening in quadrature (FWHM=7.2$\pm$0.2 \AA\  or 5.4$\pm$0.2 \AA\ depending on the observing run).  Slit effects could be present
specially during the 2008 run, since  point sources did not fill the slit   
(seeing FWHM=0.65$\pm$0.05$\arcsec$ vs. 1.3$\arcsec$ slit). 
We do not expect this to affect the conclusions presented in this paper.

\section{Analysis and results}

\begin{figure*}
\includegraphics{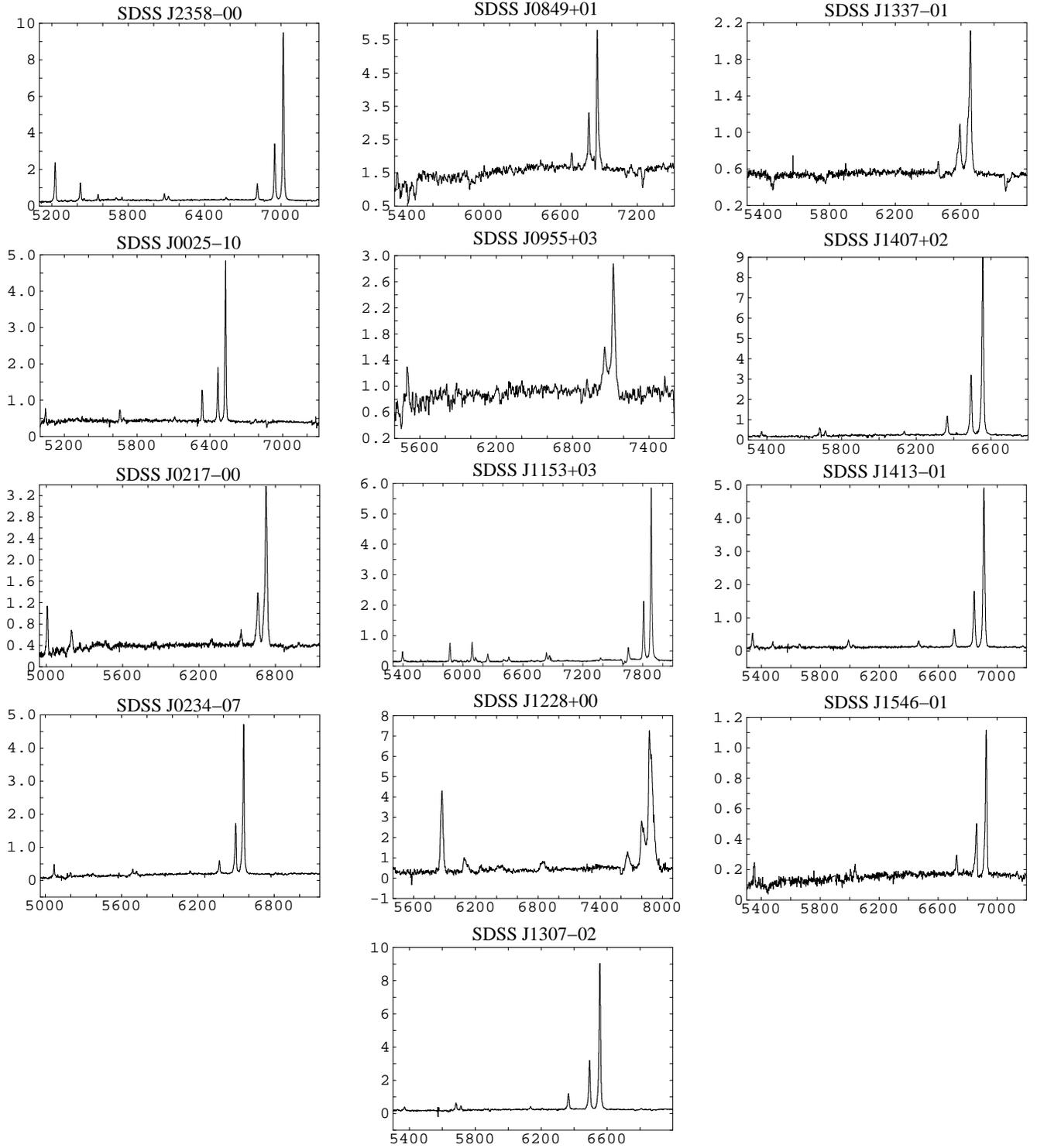}
\vspace{8in}
\caption{Nuclear spectra of the quasars studied in this work. The flux is given in units of 10$^{-16}$ erg s$^{-1}$ cm$^{-2}$ 
and the wavelength in \AA\ in all cases except for SDSS J0849+01, SDSS J0955+03, SDSS J1228+00 and SDSS J156-00 where the flux factor is 10$^{-17}$.  The strongest line in the blue is [OII]$\lambda$3727. H$\beta$ and [OIII]$\lambda\lambda$4959,5007 are clearly identified at the reddest wavelengths.}
\end{figure*}

The results of the imaging and spectroscopic analysis are presented in this section. 
\begin{table*}
\label{tab:log}      
\centering                          
\begin{tabular}{llllllllll}        
\hline                
 Object  & $z$ & Date & Filter/ & Slit  & $t_{exp}$ & Seeing & $\arcsec$/kpc & Selection \\    
    &  &  grism & PA  & sec  & FWHM('') & & criteria \\    \hline
SDSS J235818.87-000919.5 &   0.402 &  08/09/08 & FILT\_691\_55+69 ([OIII])  &  &   4$\times$200  & 0.69$\pm$0.07  & 5.34 & IS,OD \\
& &   &   GRIS\_600RI & 60         & 6$\times$700 \\
& &   & GRIS\_600RI & 83       & 1$\times$700 \\ \hline
SDSS J002531.46-104022.2 & 0.303  &   08/09/08 & FILT\_485\_37+68 ([OII])   &  &   3$\times$200  & 0.79$\pm$0.02 & 4.45 & OD \\
&  & 08/09/08  & GRIS\_600RI & 60    &    3$\times$300 &   \\
 & &   09/09/08 & GRIS\_600RI & 0   &   4$\times$700 & 0.69$\pm$0.02 & \\  \hline
SDSS J021758.19-001302.7 &  0.344  &  09/09/08 & [SII]+62 ([OIII])  &   &  4$\times$200 & 0.69$\pm$0.02 & 4.85 & PK, OD \\ 
  & & & V bess   &  &    4$\times$120 \\
  & &   & GRIS\_600RI & 116     &    4$\times$700 \\ \hline
SDSS J023411.77-074538.4 &     0.310 & 09/09/08 &  H$\alpha$+83 ([OIII])  &   &  4$\times$200 & 0.67$\pm$0.04 & 4.52 & PK \\
  & &   & GRIS\_600RI & 171      &  4$\times$700 \\ \hline \hline
SDSS J084943.82+015058.2 & 0.375 & 18/04/09 &  GRIS\_600RI & 153 & 2$\times$700 &   0.88$\pm$0.08 & 5.13 & -- \\ 
 &   & 18/04/09 &  GRIS\_600RI & 65.5 &  4$\times$700  &\\ \hline
SDSS J095514.11+034654.2 & 0.422 & 17/04/09 & GRIS\_600RI & 115 & 6$\times$700   &    1.60$\pm$0.04 & 5.52 & PK \\ \hline
SDSS J115314.36+032658.6  & 0.575 & 18/04/09 &  GRIS\_600RI &  65 &  6$\times$700 &   0.8$\pm$0.1 & 6.53 & -- \\ \hline
SDSS J122845.74+005018.7  & 0.575 & 17/04/09 & GRIS\_600RI & 170 & 6$\times$700  &    1.6$\pm$0.1 & 6.54 & PK \\ \hline
SDSS J130740.56-021455.3   & 0.425  & 18/04/09 & GRIS\_600RI & 76   & 6$\times$700 &  0.83$\pm$0.03   & 5.54 &  PK \\ \hline
SDSS J133735.02-012815.7  & 0.329 & 17/04/09 &  GRIS\_600RI & 159 & 6$\times$700 &     1.47$\pm$0.09 & 5.60 & OD,PK \\ \hline
SDSS J140740.06+021748.3  & 0.309 & 18/04/09 & GRIS\_600RI &  150 & 4$\times$700  &    0.73$\pm$0.04 & 4.52 & OD \\ \hline
SDSS J141315.31-014221.0  & 0.380 & 18/04/09 &  GRIS\_600RI & 45 & 6$\times$700 & 1.54$\pm$0.09 & 5.18 & OD\\ \hline
SDSS J154613.27-000513.5   & 0.383 & 18/04/09 & GRIS\_600RI &  109 & 6$\times$700 &   0.90$\pm$0.05 & 5.20 & --  \\ \hline
\end{tabular}
\caption{Summary of the observations. The redshift  $z$ (column 2) is the value measured with the FORS2 spectra. The filter used for the intermediate or narrow band  images
is shown in column 4 for those objects for which a continuum+emission line image exists. The  emission line  within the filter spectral window is indicated in brackets.
The GRIS$\_$600RI grism was used to obtain the spectra of all objects.
The slit position angles (PA) measured North to East are shown in column 5. The exposure time in seconds appears
  in 
column 6. The seeing FWHM values (column 7) for each object were measured with several stars in the images 
 of that specific object.  The spatial scale per arcsec at the corresponding $z$ is given in column 8. Column 9 shows specific criteria used for the selection of the objects
(see text). ``IS'', ``OD'' and ``PK'' mean ``Interacting System'', ``Other Data'' and ``Perturbed kinematics'' respectively. The 2008 and 2009 runs are separated by a double horizontal line}
\end{table*}

Several considerations must be taken into account. 
 One of the studies we perform here is the characterization of 
the gas ionization properties with the goal of exploring the nature of  the excitation mechanism  at different spatial locations for each quasar (stellar vs. AGN photoionization, i.e., ionization by the continuum  emitted by the quasar). 
The diagnostic diagram [OIII]$\lambda$5007/H$\beta$ vs. [OII]$\lambda$3727/H$\beta$ is used for this purpose.
We chose this  diagram because it involves the strongest lines detected in the spectra of most spatial locations under consideration and because they provide useful  information
about the presence of both AGN and/or stellar photoionization. 
Diagnostic diagrams will be shown only for objects with [OII]  within the observed spectral range and with detected extended emission line structures.

The data will be compared with predictions from the standard photoionization
model
sequence  (Robinson et al. \citeyear{rob87}) often applied to low and high redshift type 2 active galaxies  (see Villar-Mart\'\i n et al. \citeyear{vil08}, \citeyear{vil10} for a discussion about its application to  type 2 quasars). The models assume solar metallicity, gas density $n$=100 cm$^{-3}$ and  power-law index $\alpha$=-1.5.   Reddening is ignored. The ionization parameter $U$ varies along the sequence\footnote{$U=\frac{Q}{4~\pi~r^2~n~c}$, where Q is the quasar photon ionizing luminosity, $r$ is the radial distance  to the ionizing source, $n$  is the gas density and $c$  is the speed of light.} (models with log($U$)= -3, -2 and -1 are marked in each diagram).  
The assumption of this $n$ value is supported by measurements  of EELR densities associated with type 1 quasars in the range $\sim$several tens to several hundred cm$^{-3}$  (e.g. Fu \& Stockton \citeyear{fu08}). The existence of densities as high as 10$^6$ cm$^{-3}$ in the Narrow Line Region (NLR) are not discarded, but the large strength of the [OII]$\lambda$3727 line,  which has a critical density of  $\sim$3$\times$10$^3$ cm$^{-3}$, suggests that the line fluxes have a very high contribution of much lower density gas. In addition, varying the  gas density between several cm$^{-3}$
and $<$3000 cm$^{-3}$ is equivalent to  varying the the ionization parameter $U$, which is accounted for in the models.  

The locations of the HII galaxies from the catalogue by Terlevich et al. (\citeyear{ter91}) are also  shown for comparison.

 The interpretation of the results based on these diagnostic diagrams  must be
considered with caution, since   they do not allow an
unambiguous discrimination between stellar and AGN photoionization. Other ratios such
as [NII]$\lambda$6583/H$\alpha$ or [SII]$\lambda\lambda$6716,6731/H$\alpha$ would be needed, but they are outside the spectral range covered by our data.

  Type 2 active galaxies
in general and  type 2 quasars in particular present a large scatter in this and other diagnostic diagrams
(e.g.  Lamareille \citeyear{lam04}, Villar-Mart\'\i n et al. \citeyear{vil08}), with a substantial fraction
 overlapping with  HII galaxies.   Marked differences in the location of the spectra  (e.g. location in the AGN area vs. the HII galaxy area) have been usually interpreted as a consequence of  a range
 of gas and/or  AGN ionizing continuum properties. \cite{vil08} showed that the additional  contribution of stellar photoionization, which has generally been ignored, can be  an alternative explanation.  
  We will use other information such as the spatial morphology and line kinematic properties of the ionized gas to try to disentangle between both scenarios: stellar vs. AGN photoionization. 
 
Dust extinction has been taken into account whenever possible (Osterbrock \citeyear{ost89}), since it could have an impact on the interpretation of the line ratios.  

The effects on the  measured fluxes of some lines  due to the absorption features on the  underlying host galaxy continuum are expected to be negligible for most objects in the sample, due to the large equivalent widths of the emission lines. For those few objects where this effect could be relevant, the impact on the interpretation of the line ratios  will be discussed. We have not subtracted the underlying galaxy continuum because the uncertainty on determining the  continuum level introduces larger errors on the measured line fluxes than if the original spectra are used. The conclusions of our work are not affected by this.

As in other studies of EELRs associated with quasars and radio galaxies, we define  EELRs as spatially extended regions of ionized gas  associated with the quasars, rather  than with companion objects
(e.g. the ionized gas of a companion nucleus/galaxy/knot is not considered part of an EELR). This definition is independent of gas excitation mechanism,
 the nature and the origin of the nebulae. Thus, 
 features such as ionized tidal tails/bridges are included in the concept of EELR.  When  an extended nebula of a different  or unknown nature is found associated with a quasar  we will refer to it as the quasar ionized nebula (ionization cones, for instance).

The nuclear spectra and the nuclear line ratios for all objects are shown in Fig.1 and Tables 2 and 3 respectively. The electron temperatures have been measured using the [OIII]$\lambda$5007,4959 and [OIII]$\lambda$4363 lines (Osterbrock \citeyear{ost89}).

Hereafter, [OIII], [OII], HeII  will be used instead of [OIII]$\lambda$5007, [OII]$\lambda$3727 and HeII$\lambda$4686.

\begin{table*}
\centering
\begin{tabular}{llllllllll}
\hline
 &  0217-00 & 0217-00* & 2358-00 & 2358-00 & 0234-07 & 0234-07 & 0025-10 & 0025-10 \\
Line/H$\beta$ &  Observed & Dered. & Observed & Dered. &  Observed &  Dered.& Observed &  Dered. \\ \hline
 ~[OII]    &  1.9$\pm$0.3  &  17$\pm$11  & 2.52$\pm$0.08 &  4.7$\pm$0.2   &  N/A & N/A & N/A & N/A  \\
 ~[NeIII]$\lambda$3869  & 0.93$\pm$0.15 & 6.6$\pm$4 & 0.97$\pm$0.04 & 1.71$\pm$0.08 & 0.8$\pm$0.1 & 1.2$\pm$0.3 & 0.32$\pm$0.03 & 0.60$\pm$0.09 \\
 ~[OIII]$\lambda$4363   & 0.18$\pm$0.03  &  0.50$\pm$0.18 &  0.21$\pm$0.02 & 0.28$\pm$0.03 & 0.26$\pm$0.04 & 0.32$\pm$0.05 & 0.09$\pm$0.02 & 0.13$\pm$0.01  \\
~HeII  &  0.36$\pm$0.06  &  0.5$\pm$0.1  & 0.15$\pm$0.01 & 0.17$\pm$0.01 & 0.20$\pm$0.02  &  0.22$\pm$0.02 &  0.11$\pm$0.01 & 0.13$\pm$0.01 \\
 ~[OIII]$\lambda$5007  & 14.6$\pm$2.3  &  11$\pm$2  & 10.9$\pm$0.4 & 9.9$\pm$0.4 & 12.6$\pm$0.6 &  11.8$\pm$0.7   & 4.6$\pm$0.2 & 4.2$\pm$0.2  \\
 ~T$_4$ & 1.2$\pm$0.1 & 2.5$\pm$0.7 & 1.5$\pm$0.1 & 1.8$\pm$0.1 & 1.5$\pm$0.1 &  1.78$\pm$0.16 & 1.51$\pm$0.15 & 1.9$\pm$0.1 \\
~H$\gamma$  &  0.17$\pm$0.05  &  0.47 & 0.35 $\pm$0.02  & 0.47  &  0.38$\pm$0.04 & 0.47 &   0.34$\pm$0.02 & 0.47  \\   
~H$\delta$  & No  & No   &  0.15$\pm$0.03  &  0.23$\pm$0.05  &  0.17$\pm$0.05 & 0.23$\pm$0.08  & 0.13$\pm$0.03 & 0.21$\pm$0.05 \\   \hline
~F(H$\beta$)  & 4.5$\pm$0.6 &   & 10.6$\pm$0.3  & &5.3$\pm$0.2 & &  10.7$\pm$0.5 \\ 
\hline 
\end{tabular}
\caption{Nuclear line ratios relative to H$\beta$ for the quasars observed in the 2008 run.  The ratios corrected for dust extinction
are based on the reddening derived from H$\gamma$/H$\beta$ (case B recombination value is 0.47). The H$\beta$ flux is given in units of 10$^{-16}$ erg s$^{-1}$ cm$^{-2}$. T$_4$ is the electronic temperature in units of 10$^{4}$ K. *The reddening values estimated for   0217-00 are uncertain possibly due to underlying stellar absorption (see $\delta$3.3).}
\centering
\begin{tabular}{lllllllll}
\hline
 & 0849+01 &   0955+03 & 1153+03 &  1228+00 &  1307-02 &  1307-02      \\
Line/H$\beta$ & Observed &  Observed &    Observed &     Observed &  Observed & Dered.  \\ \hline
 ~[OII]    & N/A  &   N/A &  0.8$\pm$0.1 &  4.0$\pm$0.6  &   1.96$\pm$0.08 &  2.5$\pm$0.3   \\
 ~[NeIII]$\lambda$3869  & 1.4$\pm$0.4   & 2.0$\pm$0.8 &  0.9$\pm$0.1 &   0.9$\pm$0.1  &  0.75$\pm$0.04 &  0.9$\pm$0.1    \\
 ~[OIII]$\lambda$4363   & No  &  No & 0.35$\pm$0.04    &  0.15$\pm$0.10 & 0.15$\pm$0.01  &  0.17$\pm$0.02      \\
~HeII  & 0.37$\pm$0.08 &  No  &   0.12$\pm$0.02 &   No &    0.38$\pm$0.02  &  0.40$\pm$0.02      \\
~[OIII]$\lambda$5007  & 12$\pm$1.5  & 16$\pm$4 &   14.1$\pm$0.09 &   12$\pm$1 &   10.9$\pm$0.3  & 10.5$\pm$0.3       \\
 ~T$_4$ & N/A &   N/A & 1.7$\pm$0.1   &  1.25$\pm$0.35  &    1.30$\pm$0.05   & 1.39$\pm$0.07      \\
~H$\gamma$  & No &  No &  0.47$\pm$0.06  & 0.6$\pm$0.1 &    0.42$\pm$0.02   & 0.47       \\
~H$\delta$  & No &  No & 0.25$\pm$0.03 &  0.24$\pm$0.05 &    0.21$\pm$0.02  & 0.25$\pm$0.03     &   \\
\hline 
~F(H$\beta$)  &  0.52$\pm$0.06  & 4$\pm$1 & 9.6$\pm$1.0 & 2.8$\pm$0.3  &    4.51$\pm$0.09  & 7.0$\pm$1.5  \\ 
\hline
\end{tabular}
\centering
\begin{tabular}{llllllllllll}
\hline
 &   1337-01*  &  1407+02  &  1407+02    & 1413-01 & 1413-01 & 1546-00   \\
Line/H$\beta$ & Observed & Observed & Dered. &  Observed &  Dered.& Observed  \\ \hline
 ~[OII]  &  N/A    & N/A  &  N/A   &  N/A &  N/A   & N/A     \\
 ~[NeIII]$\lambda$3869 & N/A  &  N/A  &  N/A   &  0.61$\pm$0.03 & 0.92$\pm$0.07 & 1.1$\pm$0.2       \\
 ~[OIII]$\lambda$4363   &  0.6$\pm$0.1 &  0.25$\pm$0.01  & 0.30$\pm$0.02   & 0.14$\pm$0.02  & 0.17$\pm$0.03  & 0.7$\pm$0.1    \\
~HeII  &   0.8$\pm$0.1 & 0.18$\pm$0.01  &  0.19$\pm$0.01  &  0.33$\pm$0.1  & 0.50$\pm$0.03 &   No      \\
~[OIII]$\lambda$5007 &  25$\pm$2  &  10.9$\pm$0.4   & 10.3$\pm$0.4  &  9.6$\pm$0.4  & 9.0$\pm$0.4 &  8.3$\pm$0.9       \\
 ~T$_4$ &   1.7$\pm$0.2  & 1.64$\pm$0.07   & 1.85$\pm$0.08   & 1.33$\pm$0.08  &  1.5$\pm$0.1   &  4.5$\pm$1.0    \\
~H$\gamma$   &  No  &  0.39$\pm$0.02   &  0.47 &  0.38$\pm$0.01  &0.47  &  No     \\
~H$\delta$   &  No  &  0.19$\pm$0.04   & 0.25$\pm$0.05  & 0.19$\pm$0.03 &  0.26$\pm$0.24   & No    \\
\hline 
~F(H$\beta$) &  1.28$\pm$0.08  & 10.1$\pm$0.2  & 25$\pm$4  & 75$\pm$3 & 170$\pm$13 & 1.6$\pm$0.2       \\ 
\hline 
\end{tabular}
\caption{Nuclear line ratios relative to H$\beta$ for the objects observed in the 2009 run.  When the reddening was negligible (1153+03, 1228+00) or  uncertain (0849+01, 0955+01, 1337-01, 1546-00)  no dereddened line ratios are shown. 
 N/A means that the line was outside the spectral range
or, in the case of $T_4$, that the calculation could not be done due to the non detection of [OIII]$\lambda$4363. 
``No'' means the line was not detected or too noisy.  *The H$\beta$ flux of SDSS 1337-01 is affected by underlying stellar absorption (see $\delta$3.10).}
\end{table*}

\subsection{SDSS J2358-00 ($z=$0.402)}

\begin{figure}
\includegraphics{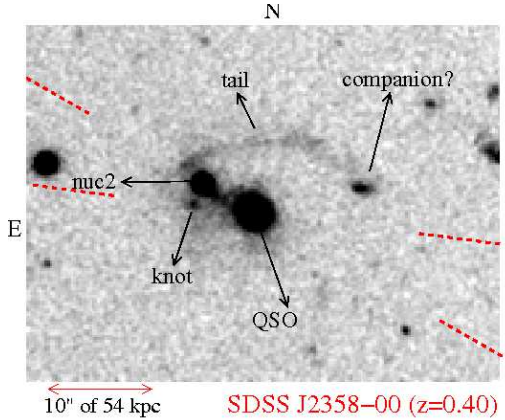}
\vspace{2.35in}
\caption{FORS2+VLT intermediate band image of  SDSS J2358-00  containing the [OIII]  line. The original image has been boxcar smoothed
with a 2x2 window. The quasar is member of an interacting system. 
A spectacular tidal tail stretches to the E apparently joining the companion nucleus ($nuc2$) with another galaxy. The 
knot identified in the figure is probably at the same $z$ (see text). The location of the PA 60 and 83 slits
used to obtain the long slit spectra are indicated with red dashed lines.}
\end{figure}

 This quasar is member of an interacting system (Fig. 2;  see also Zakamska et al. \citeyear{zak06}). A tidal
tail stretches   from the companion nucleus  ($nuc2$ in the figure) to 
the West  and apparently joins it with another companion located at  $\sim$15$\arcsec$ or 80 kpc in
projection, although the $z$ for this object is unknown.  Low surface brightness
diffuse extended emission is detected in a larger area South of the
quasar and the companion nucleus. 
A knot is located at $\sim$5.5$\arcsec$ or 30 kpc East of the quasar.  It is elongated in the E-W direction.

\begin{figure}
\includegraphics{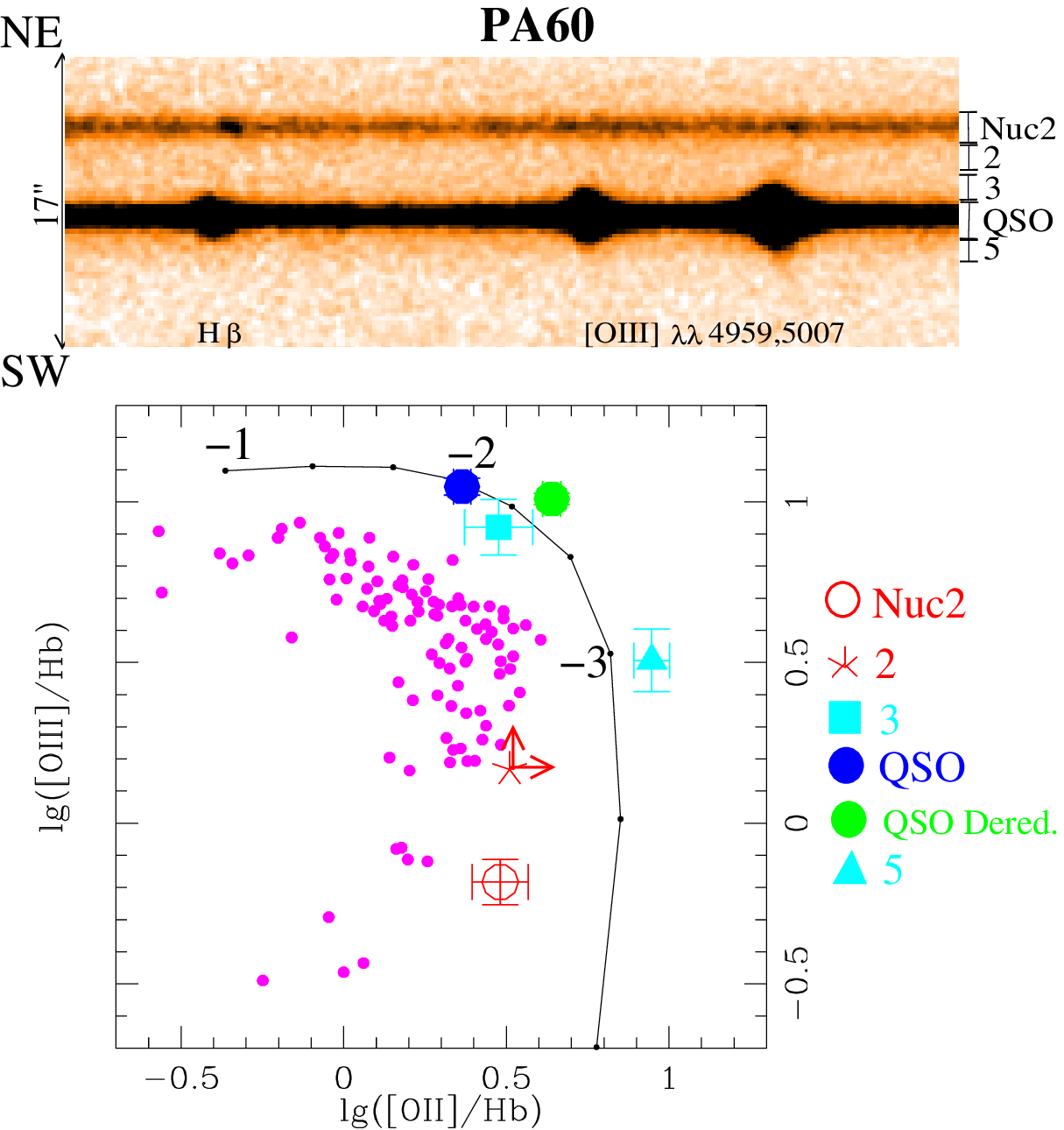}
\includegraphics{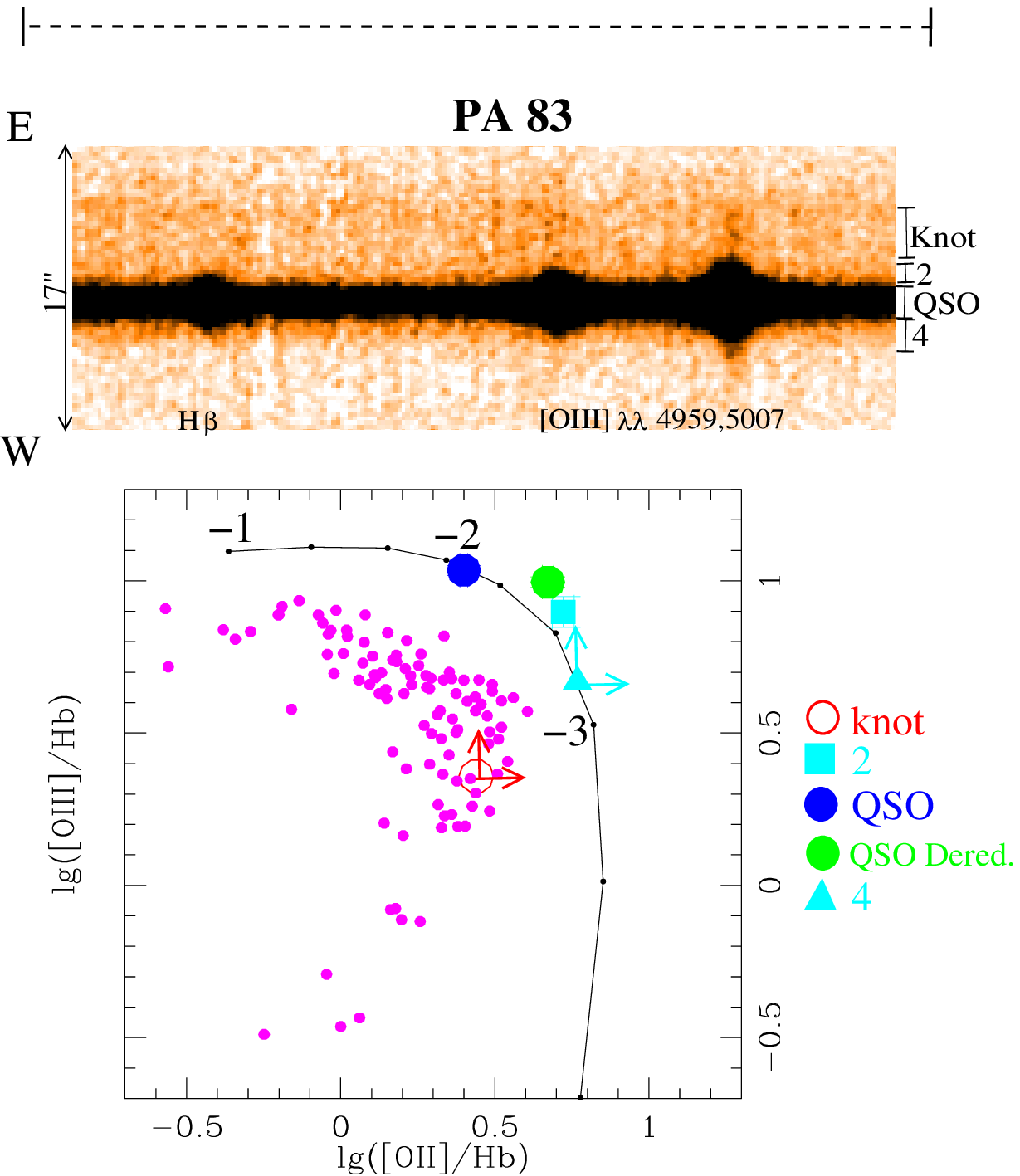}
\vspace{7.2in}
\caption{SDSS J2358-00. The H$\beta$-[OIII] 2D spectra along PA 60 and PA 83 are shown. The  apertures used to extract 1D spectra of different
spatial regions are indicated. Both spectra have been boxcar smoothed with a 2x2 window. In the diagnostic diagrams, the continuous 
black line represents the standard sequence of type 2 AGN photoionization models (see text). The ionization parameter varies along the
sequence.  log($U$) values -1, -2 and -3 are indicated. The small magenta symbols are the HII galaxies from Terlevich et al. (1991) catalogue.
 The line ratios for the individual spectra are shown on the  corresponding diagnostic diagram. The quasar nuclear line ratios corrected (green solid circle) 
and uncorrected (blue solid circle) for reddening, as well as the ratios of the adjacent regions (cyan symbols) are successfully explained by the standard AGN photoionization
models along both slit positions. The  low values of the line ratios for the companion galaxy nucleus ($nuc2$, top diagnostic diagram) are more consistent
with  HII galaxies.}
\end{figure}

The 2D spectra of the H$\beta$-[OIII] spectral region for both slit position angles (PAs) are shown in  Fig.~3. PA 60 goes through the quasar and the companion nucleus $nuc2$, which emits both continuum and strong emission lines.
PA  83 goes through the quasar nucleus and the  knot.
Low surface brightness, diffuse  extended [OIII] emission is detected at both sides of the continuum centroid along both PAs for total extensions of 7.5$\arcsec$ or 40 kpc
along PA 60 and 12$\arcsec$ or 64 kpc along PA83.

Several  apertures were used to extract 1D spectra from different spatial regions along both slits, which are also shown in Fig.~2.
The location of the line ratios measured for the individual spectra are shown in the diagnostic diagrams in the same figure.
 The quasar nuclear line ratios both corrected  (green solid circle)
and uncorrected (dark blue solid circle) for  dust reddening 
   and the ratios of the adjacent spatial regions (cyan symbols) are successfully explained by
AGN photoionization along both PAs. 

On the contrary, the spectrum of the companion ($nuc2$) shows very low line ratios, noticeably lower than those measured in typical type 2 AGNs.
Similar values
are often measured  in HII galaxies (e.g.  Lamareille et al. \citeyear{lam04}). Dust reddening could shift an AGN to the HII galaxy region in the diagram, since the [OII]/H$\beta$ would be
observed to be lower, while [OIII]/H$\beta$ would be similar. 
It is not possible to estimate the reddening from the Balmer emission lines, because the  $nuc2$ spectrum is noisy. However, the very low [OIII]/H$\beta$=0.5$\pm$0.1 is more suggestive of HII galaxies than type 2 AGNs. Moreover,
its emission lines are very  narrow (FWHM$\la$150 km  s$^{-1}$)
compared with typical values of type 2 AGNs (several hundred km  s$^{-1}$). Both properties suggest that
this is the actively star forming nucleus of a companion galaxy. The shift in velocity relative to the quasar is +300$\pm$80  km  s$^{-1}$ (this error takes into
account  slit effect uncertainties, see $\delta$2). This is within the range
expected for merging system ($\la$few hundred km   s$^{-1}$).

Both continuum and line emission  are detected at the location of the knot
and at the same $z$ as the quasar.  The spectrum  is rather noisy and H$\beta$ is not detected.
We can only say that the location in the diagnostic diagrams (Fig.~3, bottom) using the
H$\beta$ flux upper limits overlaps with the HII galaxies area and lies far
from the standard AGN sequence.

The knot looks sharper in the image than 
it does in the long-slit spectrum (see Fig.~2 and 3). This and the similar 
ratios between the fluxes of the knot and the stars  in the intermediate and the broad band images suggest that it is dominated by continuum emission. This feature appears very prominently in the rest frame V band ACS HST image  (Zakamska et al. 2006). It shows knotty internal substructure and it  seems to be part of
the complex system of knots and filaments associated with the quasar. This confirms that 
 the knot  belongs to the quasar system.

Based on these results, we propose that SDSS J2358-00 is  interacting with
a companion star forming  galaxy whose nuclear gas is photoionized by young stars  possibly formed as a consequence of the interaction. The
quasar is associated with an extended ionized nebula of maximum extension $\sim$64 kpc. The nuclear gas and 
the extended adjacent gaseous regions are preferentially photoionized by the quasar.

\subsection{SDSS J0025-10 ($z=$0.303).}

The VLT-FORS2 images of this quasar show  two nuclei (one hosts the quasar) separated in projection
by 1.1$\arcsec$ or $\sim$5 kpc  and two tidal tails extending
North and South (Fig~4). These are clear signatures of an ongoing merger.
Two   knots are also detected to the East ($knot1$ in the figure) and to the West ($knot2$).
$knot1$ appears much more prominently in the intermediate band image, which contains the [OII] line, while it is undetected in the broad band image. This  suggests $knot1$
is an emission line object at similar $z$ as the quasar. Its compact appearance  is reminiscent of a
star forming knot.
The opposite happens with $knot2$. It appears much more prominently relative to other features in the broad band image and thus it is most probably a  continuum source of unknown $z$.

\begin{figure}
\includegraphics{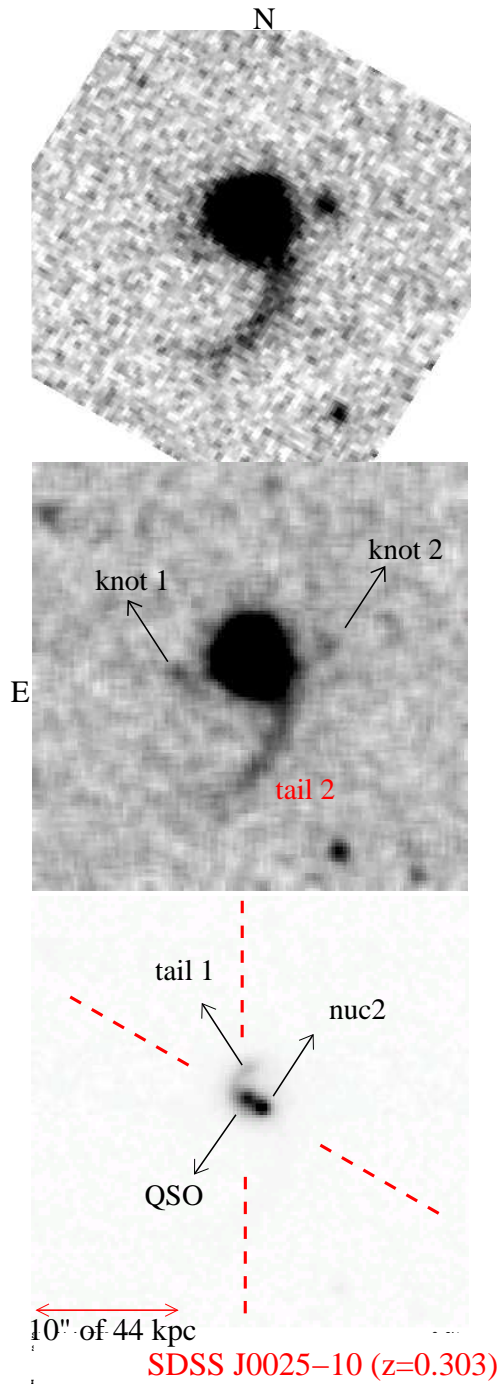}
\vspace{7.4in}
\caption{Top panel: FORS2+VLT V-Bess broad band image of SDSS J0025-10,  boxcar smoothed with a 2x2 window. Middle and bottom panels: intermediate band image
containing the [OII] line, shown  
with two different contrasts to highlight the faint (middle, boxcar smoothed with a 2x2 window) and bright (bottom, non smoothed) features.
The double nucleus and tidal tails are clear signatures of an ongoing merger. The  PA 0 and PA 60  slits are indicated with red dashed lines.}
\end{figure}

We show in Fig.~5 the H$\beta$-[OIII] 2D spectra for the two slit PAs.  
PA 0  goes through the quasar and the northern tidal tail ($tail1$) and  PA 60  goes through the quasar nucleus and the nucleus of the
companion ($nuc2$, Fig.~4). The visual inspection of both  spectra shows that the tidal tail and the companion nucleus emit strong continuum and emission lines,
which are much narrower than those of  the quasar. $nuc2$ presents a shift in velocity relative to the quasar of 0$\pm$70  km s$^{-1}$ 
(slit effects are also accounted for, see $\delta$2). The velocity blueshift of the tidal tail relative to the quasar is also small (-15$\pm$70  km s$^{-1}$).

The [OII] line was outside the observed spectral range and no  diagnostic diagram is presented.    The detection of strong HeII  (Table 2)  is consistent with values measured in type 2 AGNs, which
implies that
 AGN photoionization is the dominant mechanisms responsible for the ionization
of the nuclear gas. 
 This is further confirmed by the detection of [NeV]$\lambda$3426 in the SDSS spectrum. This, as for the rest of the sample, is to be expected since one of the selection criteria applied 
 by Zakamska et al. (\citeyear{zak03}) was an emission line spectrum typical of type 2 AGNs.

\begin{figure}
\includegraphics{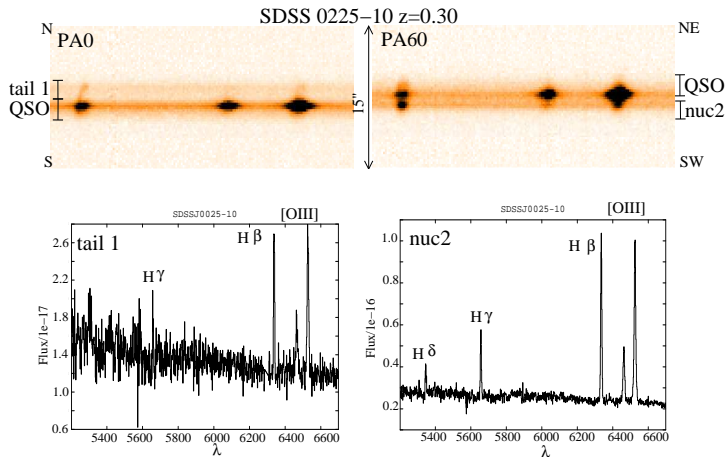}
\vspace{2.55in}
\caption{SDSS J0025-00. The spectral properties of the  the tidal tail and the companion nucleus ($nuc2$) are typical of HII galaxies/regions.} 
\end{figure}

The spectra of the companion nucleus  and the tidal tail are more
reminiscent of HII galaxies than AGNs (see Fig.~5, bottom). The
[OIII]/H$\beta$ values (1.26$\pm$0.02 and  1.17$\pm$0.08 respectively) are very low compared
with typical type 2 AGNs 
and similar to values frequently  measured in  HII galaxies  
(e.g.  Lamareille et al. \citeyear{lam04}).   HeII is undetected in $nuc2$ and the tidal tail, with 
HeII/H$\beta<$0.02  and $<$0.05 respectively. These low values are consistent with HII galaxies and smaller than ratios measured for typical type 2 AGNs (usually $>$0.1, e.g. Robinson et al. \citeyear{rob87})

The emission lines  are narrower in the tidal tail and $nuc2$ than at the quasar nucleus.  
FWHM[OIII] = 330$\pm$10   km s$^{-1}$ and  
FWHM(H$\beta$)$\leq$140  km s$^{-1}$  for $nuc2$; FWHM[OIII] = 310$\pm$30 km s$^{-1}$ and  FWHM(H$\beta$)$\le$140  km s$^{-1}$  for the tidal tail.   The quasar nucleus shows FWHM = 400$\pm$40 km s$^{-1}$ 
and 380$\pm$20  for [OIII] and H$\beta$  respectively.

Thus,  SDSS J0025-10 is undergoing a merger process with a companion star forming galaxy. We propose that both the companion nucleus and the northern 
tidal tail are photoionized by young stars which have probably formed as a consequence of the interaction process.  $knot1$
is  probably a companion star forming object.
  We find no clear evidence for a quasar extended ionized nebula  along PA 0 or PA 60
at surface brightness levels $\ga$3$\sigma$=3.2$\times$10$^{-18}$ erg s$^{-1}$ cm$^{-2}$ arcsec$^{-2}$.

\subsection{SDSS J0217-00  ($z=$0.344)}

 The VLT-FORS2   broad  and narrow band images 
are shown in Fig.~6. 
An extended diffuse structure is detected towards the N-W, whose  morphology appears more clearly defined in the broad  band image and is reminiscent of a tidal tail. A compact knot is also detected to the E, which appears relatively stronger in the narrow band image. This suggests that it is a strong line emitter at the same $z$ as the quasar.

The PA 116 slit crosses both the tidal tail and the knot.  The  H$\beta$-[OIII] 2D spectrum (Fig.~7)  shows
that the compact knot  emits strong lines and 
very faint continuum, as expected from the  images.  
Low surface brightness lines are also detected between the quasar and the knot, 
possibly emitted by  an EELR
associated with the quasar.  The  structure 
reminiscent of a tidal tail mentioned above emits only faint, diffuse continuum. 

\begin{figure}
\includegraphics{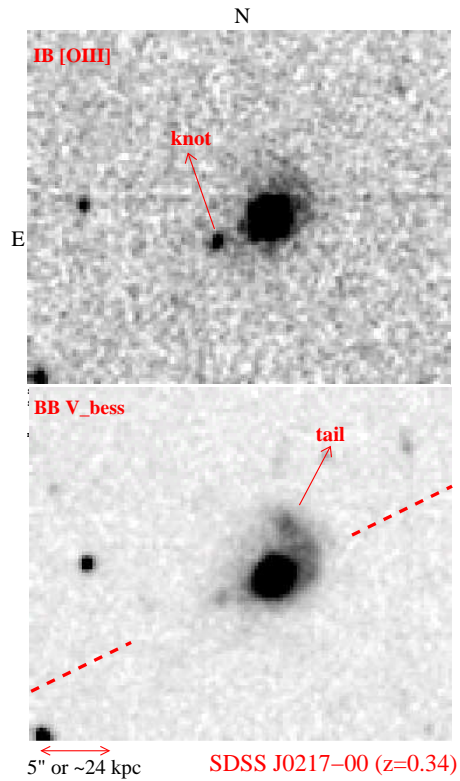}
\vspace{4.2in}
\caption{FORS2+VLT images of SDSS J0217-00.
Top: narrow band image containing the [OIII] line. Bottom: V-bess broad band image. Both have
been boxcar smoothed with a 2x2 window. Different
features appear with different prominence in the two images. The knot is much
clearer in the intermediate band image, suggesting it is dominated by line emission, while the extended diffuse structure towards the NW is much more prominent in the broad band image suggesting it emits mostly continuum. This appears like a clear tidal
tail in the broad band image. The location of the PA 116  slit  is shown on the bottom panel.}
\end{figure}

The nuclear line ratios are shown in Table 2. The reddening derived from
the Balmer lines is most likely wrong, since  the galaxy continuum is  strong
and  H$\gamma$ and H$\delta$ (not so much H$\beta$, which has larger equivalent width) are likely to be strongly absorbed. 
Given the large uncertainties due to this effect, we will 
ignore nuclear dust reddening  in the diagnostic diagram in this case.
The knot  is dominated by line emission and is not affected by this problem:
 H$\gamma$/H$\beta$=0.46$\pm$0.03 is consistent with the case B
recombination value 0.47 (Osterbrock \citeyear{ost89}). It is therefore not reddened.

Three  apertures were used to extract 1D spectra from different spatial regions along the slit (Fig.~7, top): the knot, the quasar and the intermediate region between them. 
The location of the individual spectra are shown in the diagnostic diagram.  While the nuclear spectrum
lies very far from the HII galaxies and is consistent with the AGN models (also HeII is strong, HeII/H$\beta$=0.36$\pm$0.06; Table 2),
the knot 
overlaps with the HII galaxy region.  
For the  intermediate region,   lower limits are shown,
due to the non detection of  H$\beta$. The location overlaps with the HII galaxy region as well.

The lines emitted by the knot  are split into two kinematics 
components (see Fig.~7, top). The dominant component is very narrow (FWHM $\la$120  km s$^{-1}$) and shows very low [OIII]/H$\beta$=3.6$\pm$0.3.
These properties and its location on the HII region area in the diagnostic diagrams suggest that the knot is a star forming object
where the gas is photoionized by young stars. It is redshifted by -260$\pm$70 km s$^{-1}$ relative to the quasar (uncertainties due to slit effects have been taken into account).
The faintest kinematic component is broader (FWHM=260$\pm$70) and is more highly ionized ([OIII]/H$\beta$=7.0$\pm$0.4). It is possible that this is the extension of  
 the intermediate ionized region detected between the quasar and the knot, which, as proposed above,  could be an EELR associated with the quasar. This is supported by the spatial continuity observed for the [OIII] line emission between this region and  the 
fainter kinematic component of the knot along the slit.

Therefore, the morphology of SDSS J0217-00 is reminiscent  of a galaxy with a continuum tidal tail which is a signature of an merger event.
The quasar is associated with a  companion star forming knot where the gas is  photoionized by the young stars. 
An EELR is detected across $\sim$4$\arcsec$ or 19 kpc between both objects and it might  connect them physically. 

\begin{figure}
\includegraphics{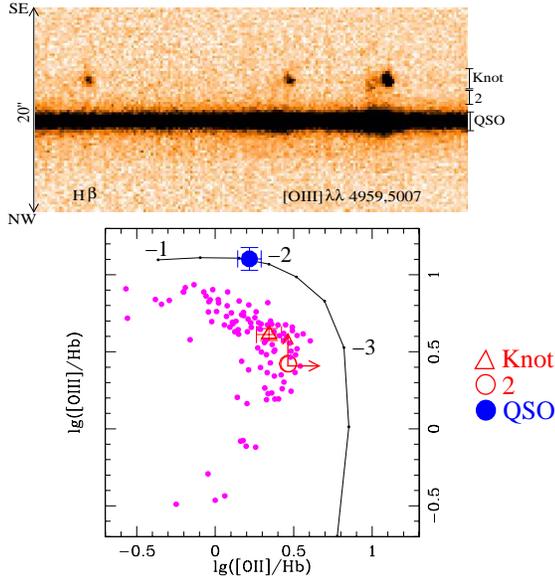}
\vspace{3.1in}
\caption{H$\beta$-[OIII] 2D spectrum of SDSS J0217-00 (top). The  apertures used to extract 1D spectra of different
spatial regions are indicated. The line ratios for the individual spectra are shown in the diagnostic diagram below. The quasar nuclear line ratios
(blue solid circle)  are successfully explained by the standard AGN photoionization
models. The companion knot (open red triangle) overlaps with the region occupied by HII galaxies. Other symbols and lines as in Fig.~3.}
\end{figure}

\subsection{SDSS0234-07 ($z=$0.310)}

No extended structures are detected in the broad or the narrow band images (Fig.~8) of this quasar. The faint object, marked as G1 in the figure
falls within the slit. The spectrum shows that it is a faint continuum source of unknown $z$.

The spatial profile of the [OIII] line (Fig.~9) is dominated by a spatially unresolved component of FWHM  0.57$\pm$0.05$\arcsec$  (vs. 0.67$\pm$0.04$\arcsec$  FWHM seeing measured on the images).
There is no evidence for extended line emission along PA 171 associated with this quasar at surface brightness levels $\ga$3$\sigma$=3.5$\times$10$^{-18}$ erg s$^{-1}$ cm$^{-2}$ arcsec$^{-2}$.

\begin{figure}
\includegraphics{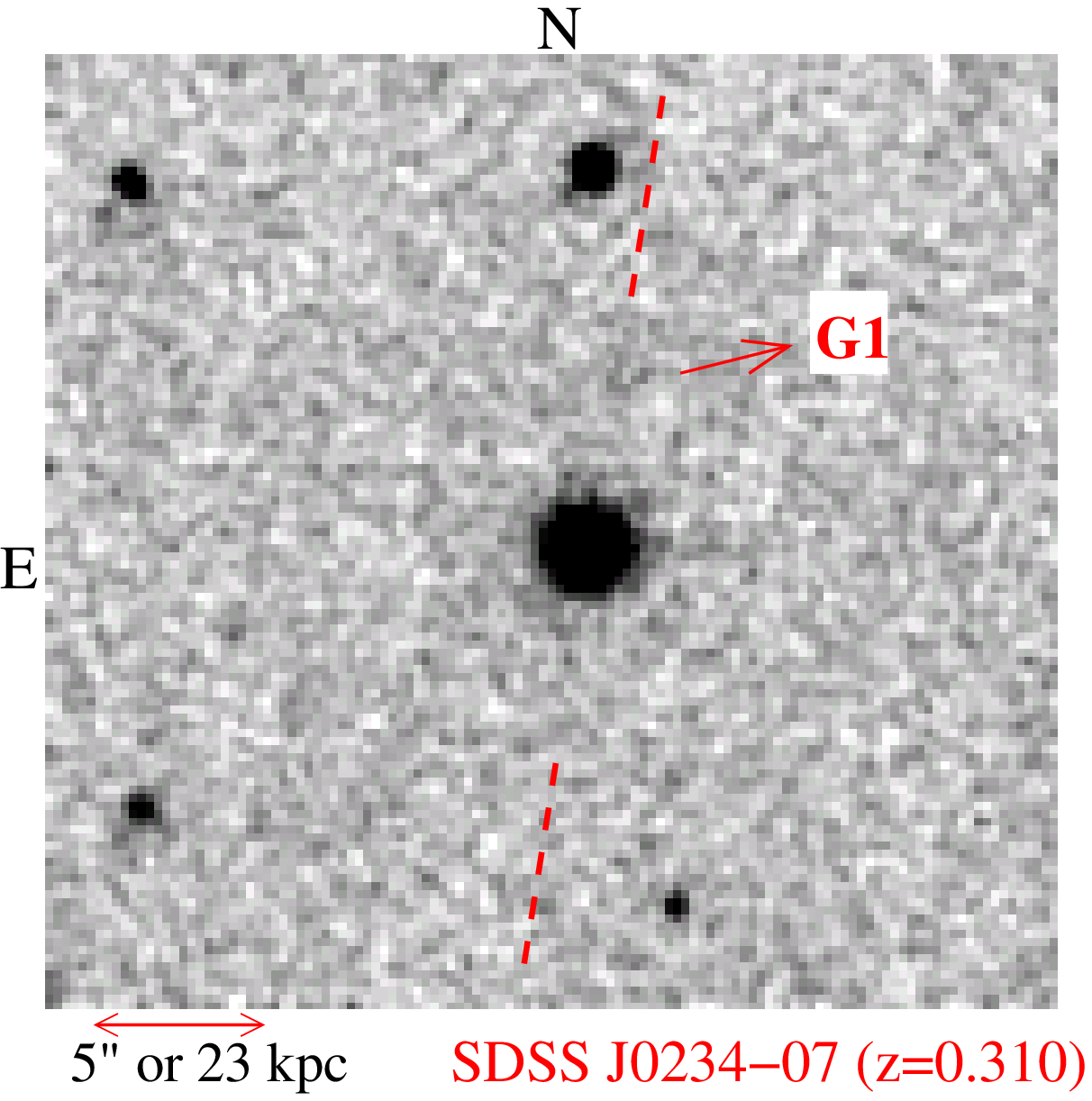}
\vspace{2.4in}
\caption{Narrow band image of SDSS J0234-00 containing   [OIII].  The original image has been boxcar smoothed with
a 2x2 window.
The spectrum shows that G1 is a continuum source of unknown $z$.}
\includegraphics{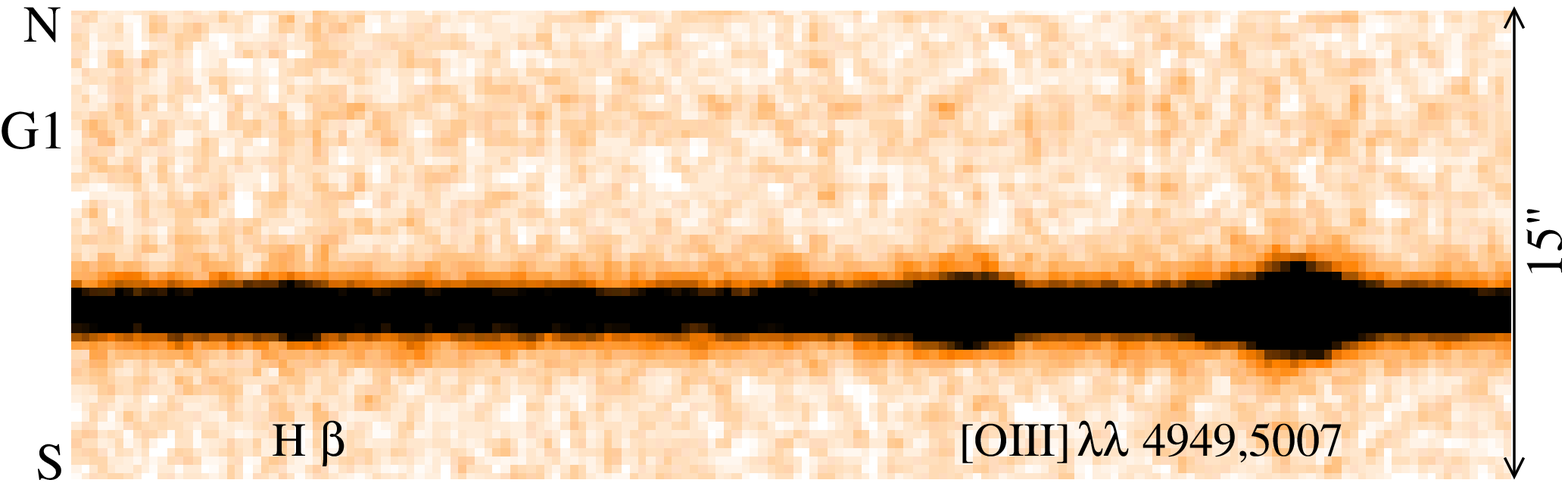}
\vspace{1.5in}
\caption{H$\beta$-[OIII] 2D spectrum of SDSS J0234-07. The original spectrum has been boxcar smoothed with
a 2x2 window There is no evidence for extended line emission along PA 171.}
\end{figure}

\vspace{0.5cm}

The objects discussed below were observed during the 2009 run. As mentioned in $\delta$2, no narrow or intermediate emission line+continuum images are available.
Therefore, except in two cases where the continuum dominated images revealed or suggested a galaxy interaction  or a peculiar quasar morphology, the slit was placed blindly or through
objects near the quasar, with the goal of  checking their $z$.

\subsection{SDSS J0849+01 ($z=$0.376)}

The narrow band continuum image is shown in Fig.~10. The emission line galaxies G1 and G2  fall within the slits. G2 lies at $z$=0.221. 
G1 emits a single emission line whose most likely identification is [OII]$\lambda$3727 at $z$=0.950 (rather than Ly$\alpha$ at $z=$4.98), given the strong continuum emission bluewards of
the line. 

The spatial distribution of [OIII] (Fig. 11) along PA 65    is dominated by
a  compact component  which is successfully represented by a Gaussian function of FWHM=1.05$\pm$0.05$\arcsec$ (vs. seeing FWHM=0.88$\pm$0.08). Taking errors into account and possible seeing variations during the spectroscopic observations
it is not possible to discern whether the [OIII] profile along PA65  is spatially resolved.
The  emission lines are unresolved along PA 153 (FWHM=0.82$\pm$0.04$\arcsec$).

\begin{figure}
\includegraphics{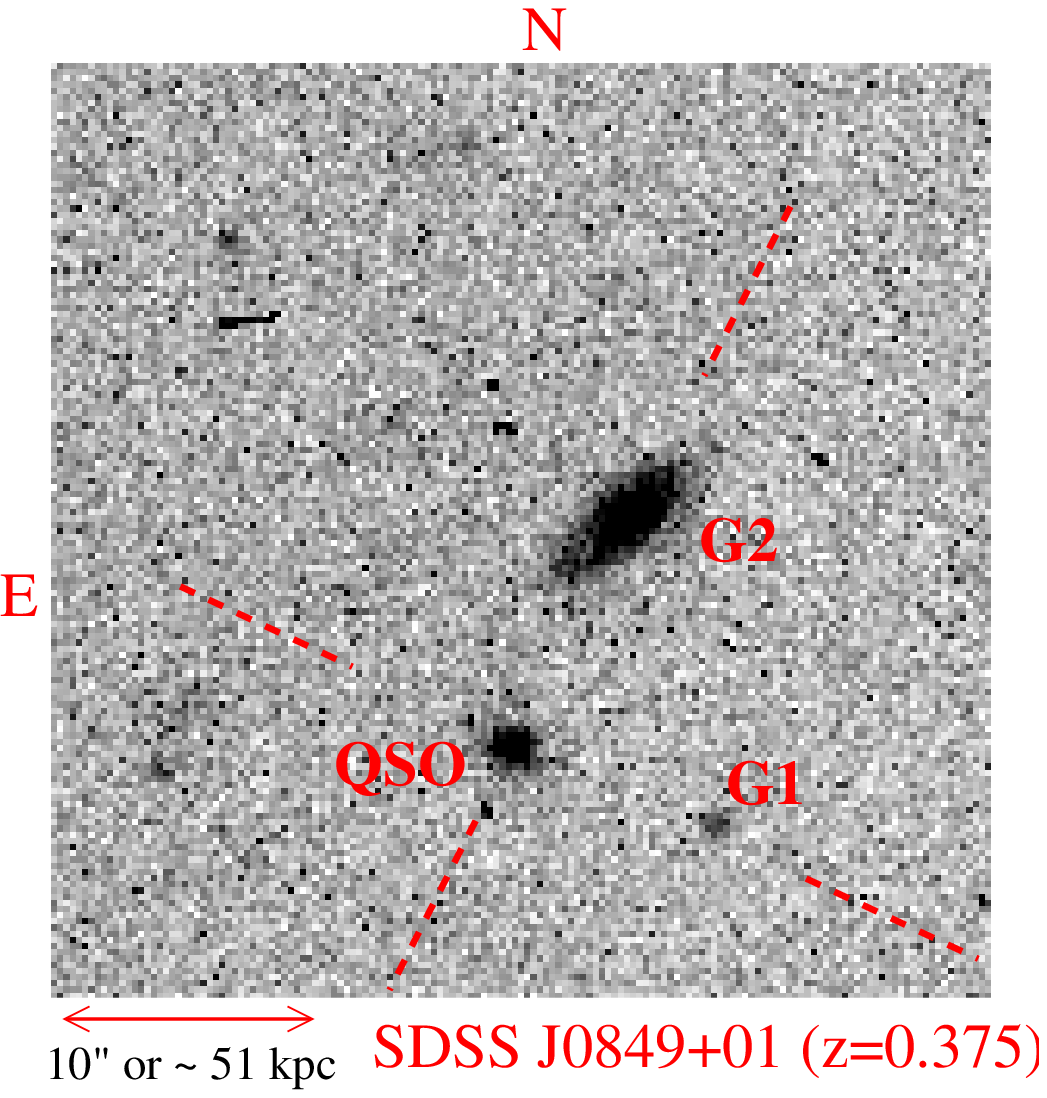}
\vspace{2.5in}
\caption{Narrow band continuum image of SDSS J0849+01 (identified with QSO).  It was obtained with the HeII6500+49 FORS2 filter and covers the rest frame spectral range $\sim$3450-3500 \AA. The slit PAs 153 and 65  were selected to
include galaxies  G1 and G2. Both are emission line galaxies at $z$=0.950 and 0.221 respectively. }
\includegraphics{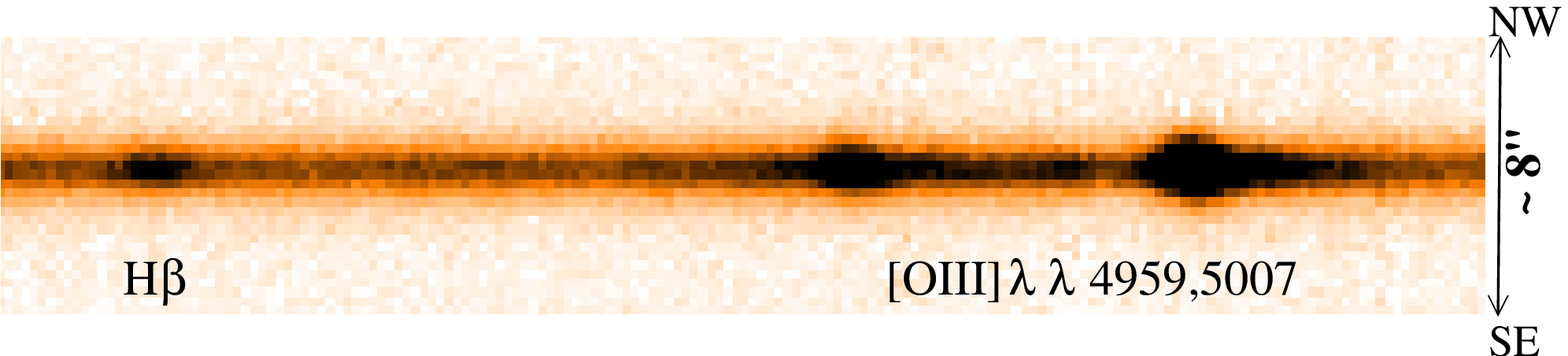}
\vspace{1.2in}
\caption{H$\beta$-[OIII] 2D spectrum of SDSS J0849+01.  G1 (Fig. 12) is also
detected. The faint line emission coinciding with H$\beta$ confirms it is at the same $z$ as the quasar.}
\end{figure}

\subsection{SDSS J0955+03 ($z=$0.422)}

The   broad band image is shown in Fig.~12.  Several galaxies marked as G1, G2, G3, G4 and G5 in Fig.~9 fall 
within the slit. The spectra reveal that  they are all  emission line galaxies  at different
$z$ than the quasar.

The spatial profile of the  emission lines along PA 115 (Fig.~13) is  dominated by a barely
resolved central component of FWHM=1.86$\pm$0.04$\arcsec$ (vs. 1.60$\pm$0.04$\arcsec$).  
The seeing FWHM was very similar at the beginning and the end of this
object exposures, so seeing variations during the spectroscopic observations are not likely to play a role and the line emission is actually spatially extended. The implied 
intrinsic size is $\sim$1$\arcsec$ or $\sim$5.5 kpc. 
Given the strong contamination by  the nuclear emission, it is not possible to isolate the emission from this
EELR to analyse the kinematic and ionization properties. In addition, very faint emission is  detected at $\sim$3$\sigma$ level  
towards the West (indicated with ``EELR'' in Fig.~13) up to a radial extent  of 7$\arcsec$ or 38.5 kpc from the continuum centroid.

\begin{figure}
\includegraphics{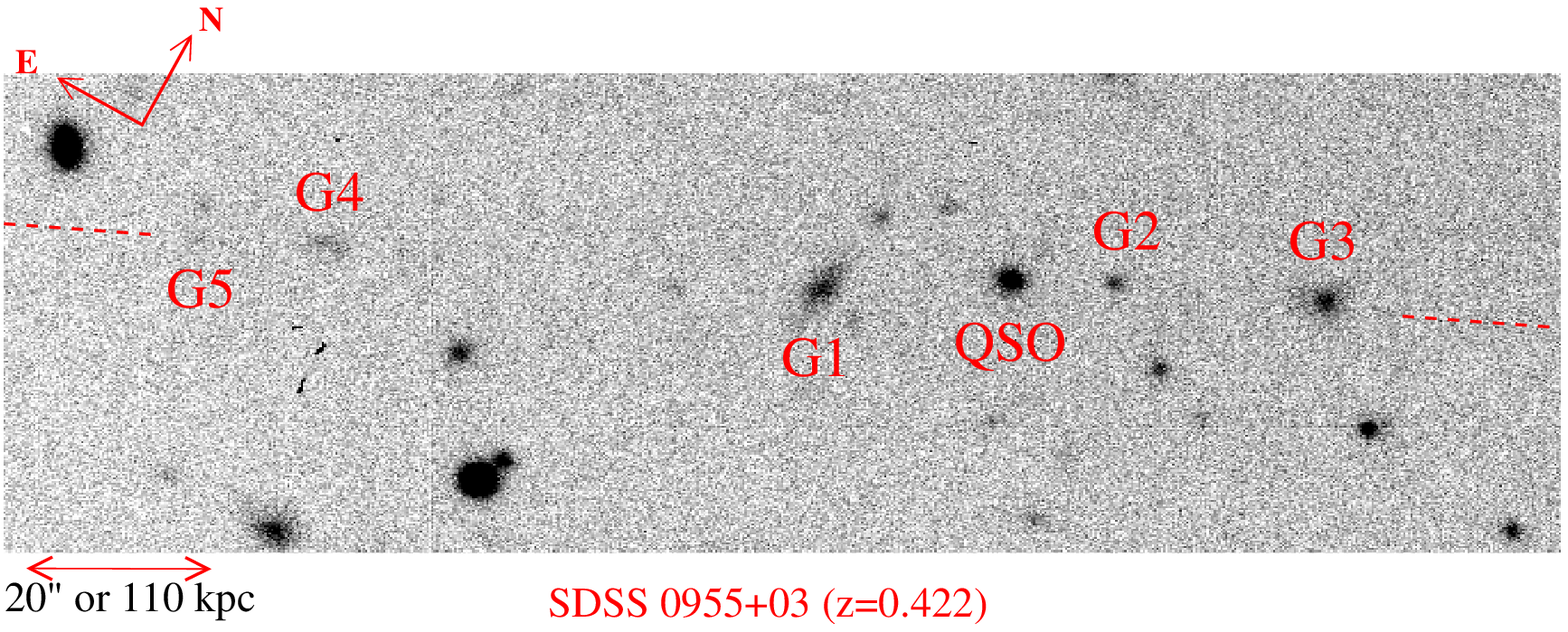}
\vspace{1.5in}
\caption{Broad band  image of SDSS J0955+03 (QSO in the image). It was obtained with the V\_High FORS2 filter
and covers the rest frame spectral range $\sim$3470-4335 \AA. 
 The slit  was
placed at PA 115, as marked by the dashed red line.  G1, G2, G3, G4, G5
fall within the slit. The spectra reveal they all are emission line galaxies at  different $z$ than the quasar.}
\includegraphics{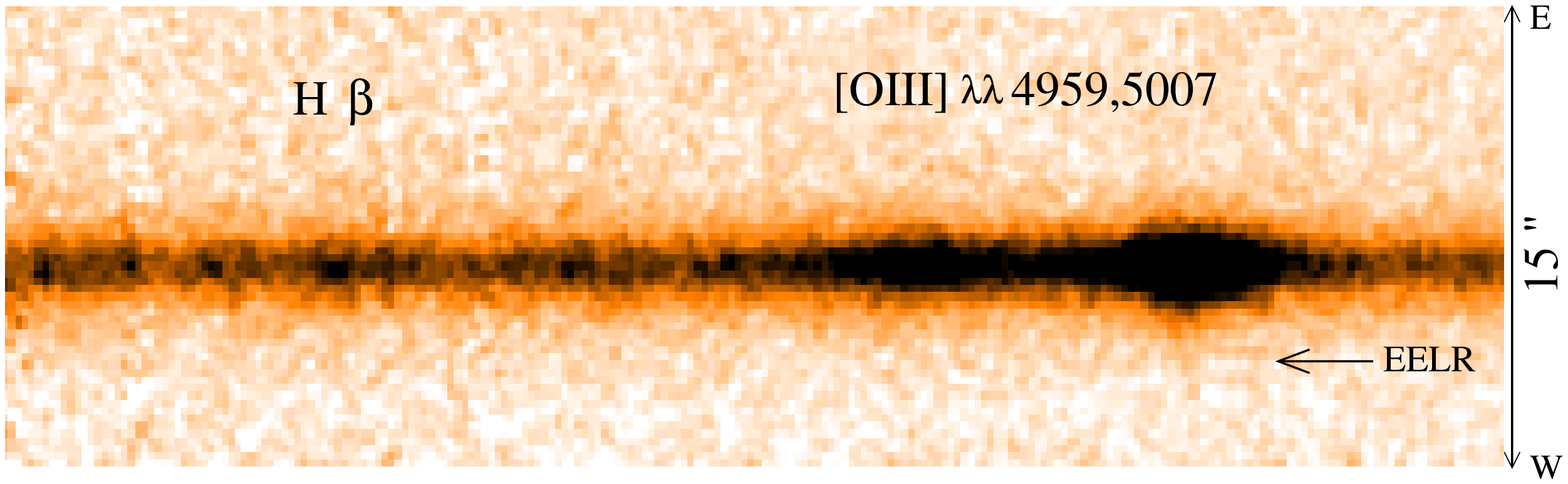}
\vspace{1.5in}
\caption{H$\beta$-[OIII] 2D spectrum of SDSS J0955+03.  The original spectrum
has been boxcar smoothed with a 2x2 window. Very faint extended emission is  detected towards the West (indicated
with ``EELR'' in the figure.)}
\end{figure}

\subsection{SDSS J1153+03 ($z=$0.575)}

\begin{figure}
\includegraphics{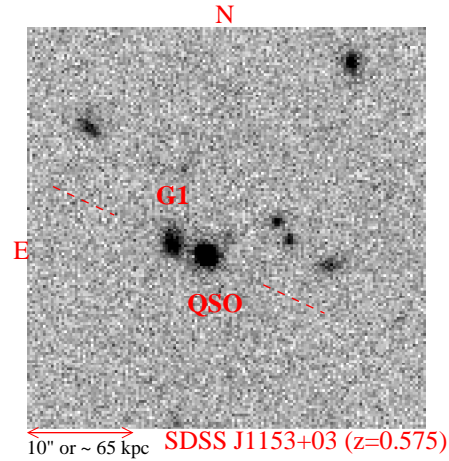}
\vspace{2.3in}
\caption{Narrow band continuum image of  SDSS J1153+03 (identified with QSO). It was obtained with the H$\alpha$+83 FORS2 filter and it covers
the rest frame spectral range $\sim$4140-4185 \AA. The  PA 65 slit is indicated. 
Galaxy G1 is at the same $z$ as the quasar and they are probably  interacting.}
\end{figure}

This quasar (Fig.â14) appears to  be interacting with a companion galaxy
(G1 in the figure), although a chance projection cannot be discarded from the image alone. The FORS2 slit was located at PA 65, crossing both nuclei.

The nuclear spectrum shows strong continuum compared with the other quasars in the sample and a very broad underlying H$\beta$ component (Fig.â15). The flux of this component  has not been included in the H$\beta$ flux used to calculate the nuclear line ratios (Table 2).  
Broad underlying HeII might also be present. Several [FeVII] emission lines are detected ([FeVII]$\lambda$3586, [FeVII]$\lambda$3759, [FeVII]$\lambda$5159).  Forbidden high ionization lines (FHIL) have been detected in the spectra of many active galaxies (e.g. Penston et al. \citeyear{pen84}, Mullaney et al. \citeyear{mul09}). Their ionization potential  is $\ga$100 eV (99 eV for Fe$^{+5}$).  Different works have suggested that they are emitted in an intermediate region 
between the NLR and the BLR (see Mullaney et al. \citeyear{mul09} for a review). More specifically, these authors propose that they are emitted by the illuminated face of the torus, a region with densities in the range $n\sim$10$^5$-10$^{10}$ cm$^{-3}$. In SDSS J1153+03, as in other AGNs, the [FeVII] lines are blueshifted relative to lower ionization forbidden lines such as [OIII] and [OII] by $\sim$200  km s$^{-1}$ and  [FeVII]$\lambda$5159, the strongest line, is much broader than all other emission lines ($\sim$1200 km s$^{-1}$ vs. several hundred km s$^{-1}$). The detection of strong continuum and very broad underlying H$\beta$ emission  in the nuclear spectrum implies a viewing angle that allows to see not only the illuminated face of the torus, but also part of the broad line and continuum regions.

\begin{figure}
\includegraphics{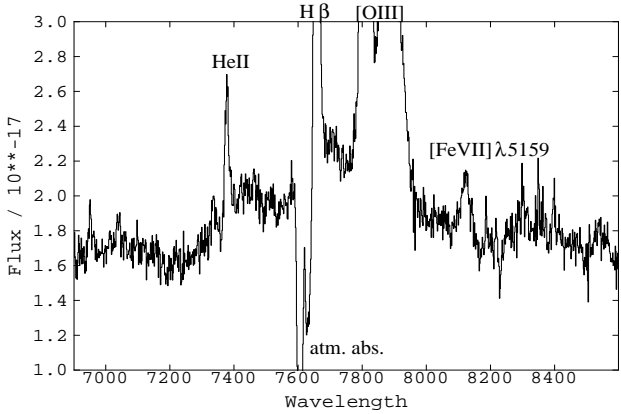}
\vspace{2.4in}
\caption{Very broad H$\beta$ is detected underneath the narrow emission lines (possibly very broad HeII  as well) in the nuclear spectrum of SDSS 1153+03.
The deep absorption feature is produced by the atmosphere. [FeVII] lines are also detected.}
\end{figure}

  Fig.~16 shows the  H$\beta$-[OIII]  2D spectrum . The quasar emission lines are clearly extended.  [OIII] 
extends across $\sim$8$\arcsec$ or 52 kpc. Line emission is detected at the location of galaxy G1 as well, but a more thorough analysis is required
to disentangle whether the lines are emitted preferentially by the quasar EELR or  G1. This is important, since one scenario or another
will determine whether G1 is at similar $z$ as the quasar.

 The continuous and rather symmetric spatial distribution of the [OIII] lines at both sides of the quasar
suggests that they are preferentially emitted by the quasar associated EELR. On the contrary, H$\beta$ and [OII] present a discontinuous, asymmetric spatial distribution, with a compact
appearance at the G1 location (see Fig.~15).  This suggests that these two low ionization lines are emitted preferentially by G1, while [OIII] has a relatively stronger contribution 
from the quasar EELR.

\begin{figure}
\includegraphics{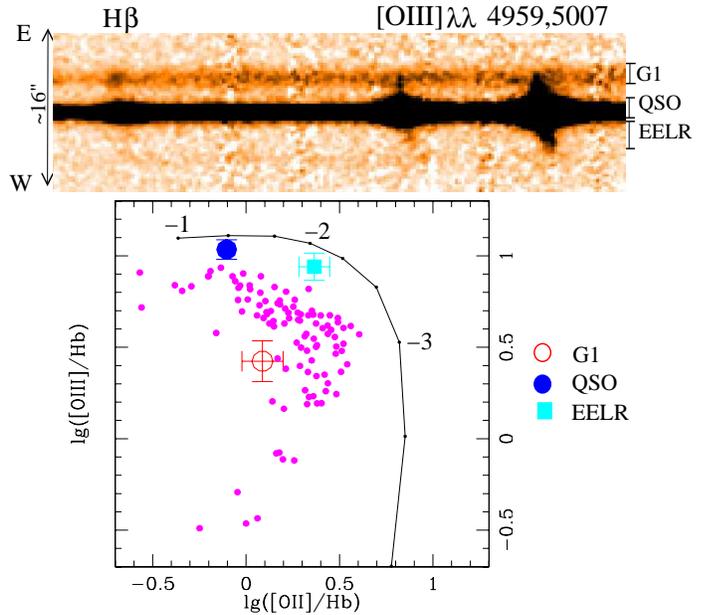}
\vspace{3.4in}
\caption{H$\beta$-[OIII]  2D spectrum of SDSS J1153+03 along PA 65 (top). 
The apertures used to extract the spectra of galaxy G1, the western EELR and the quasar are indicated. The spectrum has been  boxcar smoothed with a 2x2 window.  Notice the compact appearance
of H$\beta$ at the spatial location of G1, compared with [OIII],  which suggests that  at this location the line is excited mostly locally, rather than by the quasar (see text). 
The  diagnostic diagram shows that the nuclear line ratios (blue solid circle) and those of the Eastern EELR (cyan solid square) are successfully explained by AGN photoionization.  G1 (open red circle) overlaps with the HII galaxy region and this suggests that the gas is  photoionized by young stars in this galaxy. Other symbols and lines as in Fig.~3. }
\end{figure}

The line widths add  further information.
 [OIII] has similar FWHM at the location of G1 (280$\pm$30 km s$^{-1}$) and the western EELR (260$\pm$30 km s$^{-1}$), while  the low ionization [OII] line is, on the contrary, considerable narrower with FWHM = 200$\pm$50 km s$^{-1}$ for G1. The intrinsic width is actually narrower, since we have not taken into account that [OII]  is a doublet.   The values at  the Western EELR are 300$\pm$20  km s$^{-1}$ and  450$\pm$30 km s$^{-1}$ for H$\beta$  and [OII] respectively.  Unfortunately, the blue wing of H$\beta$ for G1 lies at the sharp edge of a deep atmospheric absorption band, so the FWHM=140   km s$^{-1}$ is highly uncertain (this does not affect
the Western EELR because of its slightly higher $z$).

The results described above  can be explained if  [OII] and H$\beta$ are emitted preferentially by G1, 
while [OIII] has a relatively large contribution of flux from the quasar EELR.  
This implies that G1 is  at the same $z$ as the quasar.   The nuclei of both objects are separated by $\sim$3$\arcsec$ or 20 kpc in projection. G1 is blueshifted
in velocity by -110$\pm$10 km  s$^{-1}$ relative to the quasar.

\begin{figure}
\includegraphics{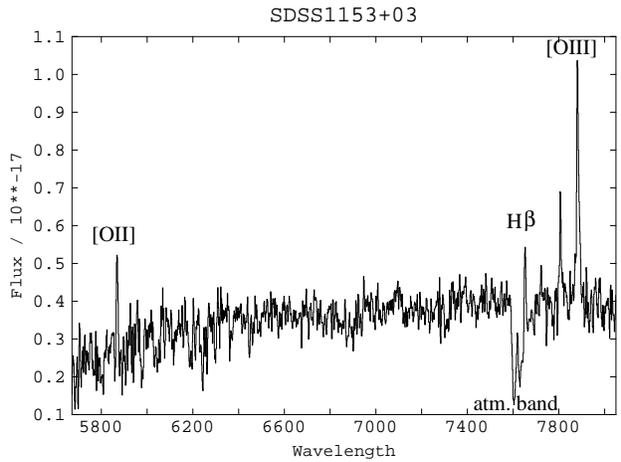}
\vspace{2.4in}
\caption{Spectrum of the companion galaxy G1 of SDSS J1153+03.  H$\beta$ is absorbed by an atmospheric absorption band.}
\end{figure}

We have extracted 1D spectra for G1 (see Fig.~17), the quasar and the western EELR.  The apertures are shown in Fig.~16 (top). The location 
in  the diagnostic diagram is shown
in the same figure (bottom). The H$\beta$ flux of the quasar is affected by atmospheric absorption. We have corrected for this effect taking into account that H$\delta$/H$\gamma$=0.53$\pm$0.02  is consistent with no reddening.  The real H$\beta$ flux is estimated assuming H$\gamma$/H$\beta$ = 0.47, the case B recombination value. This gives H$\beta$ = (9.6$\pm$0.9)$\times$10$^{-16}$ erg s$^{-1}$ cm$^{-2}$, which is the value
we have used in the diagnostic diagram. 
As expected, the quasar nuclear line ratios 
 lie very far from the HII galaxy region and are consistent with the AGN model sequence (it also emits HeII, Table 2). 
 
 The spectrum at the location of  G1 
overlaps with the region occupied by HII galaxies. It could not be corrected for line reddening, since H$\gamma$ and H$\delta$ are not detected. As we mentioned above, line reddening could
move an AGN to the HII galaxy region in the [OIII]/H$\beta$ vs. [OII]/H$\beta$ diagram. However, the atmospheric absorption works in the opposite way. 
Looking at the spectrum, we estimate that at least 50\% of the H$\beta$ flux has been absorbed. Correcting for this
[OIII]/H$\beta$ would be $\sim$1.5 , even lower than already observed and more typical of HII galaxies than AGNs. 

 The spectrum of the quasar Western EELR is consistent with AGN photoionization. Reddening would move it  even further away from the HII galaxies.

We propose that SDSS J1153+03 is interacting with
a companion star forming galaxy, where the gas is at least partially photoionized by young stars, which have probably been formed as a consequence of the interaction. 
The quasar is also associated with an ionized nebula which extends for 8$\arcsec$ or $\sim$52 kpc along PA 65.

\subsection{SDSS J1228+00 ($z=$0.575)}

\begin{figure}
\includegraphics{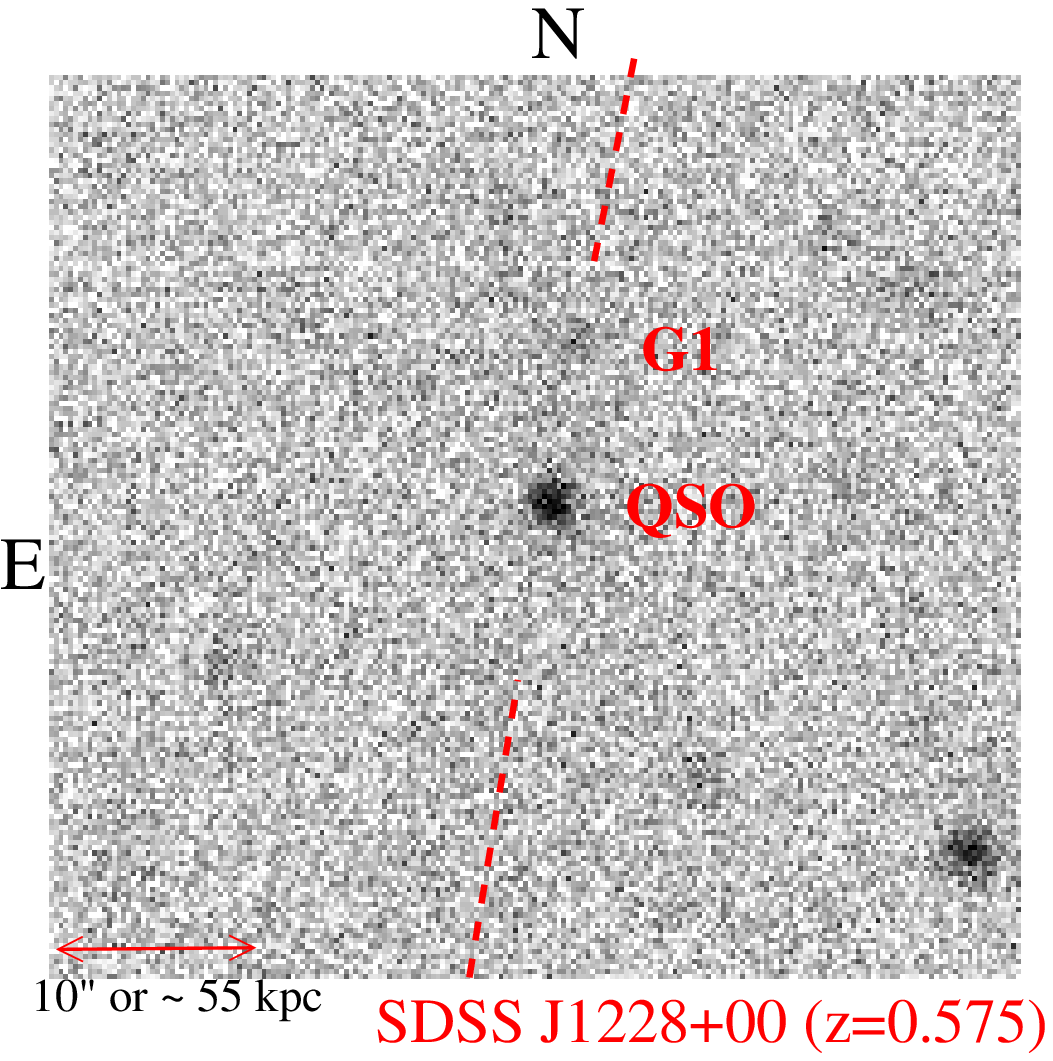}
\vspace{2.4in}
\caption{Narrow band continuum image of the field around SDSS J1228+00. It was obtained with the H$\alpha$+83  FORS2 filter 
and covers the rest frame spectral range $\sim$4150-4185 \AA.
The PA 170 slit location is indicated.  G1 is a continuum source of unknown $z$.}
\includegraphics{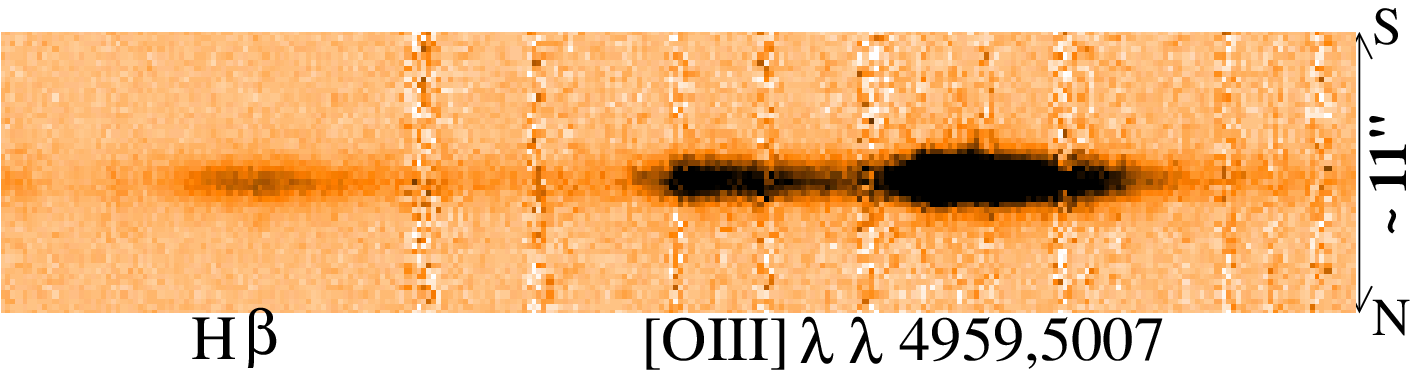}
\vspace{1.5in}
\caption{H$\beta$-[OIII] 2D spectrum  of SDSS J1228+00. No extended line emission is detected.}
\end{figure}

The narrow band continuum image  is shown in Fig.~18  and the 2D  H$\beta$-[OIII] spectrum  appears in Fig. 19.  
 The spatial profile of the [OIII] line is  unresolved  (FWHM=1.56$\pm$0.04$\arcsec$ vs. FWHM=1.6$\pm$0.1$\arcsec$ seeing)
along this direction. There is no evidence for extended line emission at surface brightness SB$\ga$3$\sigma$=3$\times$10$^{-18}$ erg s$^{-1}$ cm$^{-2}$ arcsec$^{-2}$
(Fig~19). 
G1 is a continuum source of unknown redshift.

The nuclear spectrum of this quasar (Fig.~1) is quite particular, since it shows very broad lines with FWHM$\ga$1400 km s$^{-1}$. 

This is a radio intermediate quasar (see $\delta$2).
It is known that the radio structures are capable of inducing  strong kinematic perturbation in the gas, not only in radio loud (e.g. Villar-Mart\'\i n et al.   \citeyear{villar99}), but also radio quiet active galaxies (e.g. Axon et al. \citeyear{axon98}). 
However, a more plausible scenario is that we are looking close to the edge of the obscuring torus, so that we detect emission from an
intermediate density region (10$^4$$<n<$10$^6$  cm$^{-3}$) between the broad and the narrow line regions where the lines would also have intermediate FWHM.  This is suggested by the difference in width between
 forbidden lines with high and low
critical densities ($n_{crit}$). We measure FWHM([OIII])=2000$\pm$200 km s$^{-1}$, FWHM[NeIII]$\sim$2200$\pm$200 km s$^{-1}$ and FWHM[OII]=1440$\pm$60 km s$^{-1}$. The critical densities of these lines are $n_{crit}\sim$10$^6$  cm$^{-3}$ for [OIII], $\sim$10$^7$  cm$^{-3}$  for [NeIII] and  $\sim$3$\times$10$^3$  cm$^{-3}$ for [OII]. Therefore, [OII] would be very efficiently quenched in the intermediate density region and it is
more efficiently produced in  lower density gas where the motions are also slower.
The large [OIII]/H$\beta$ ratio 12$\pm$1 suggests that the intermediate density region is very highly ionized. This  could be part of the FHIL region discussed in $\delta$3.7.

\subsection{SDSS J1307-02 ($z=$0.310)}

The  morphology of this quasar   in the broad band image (Fig.~20) is asymmetric and elongated  in the E-W direction. Faint and diffuse  emission
is detected well beyond the optical size
of the host galaxy with a  measured total extension of $\sim$15$\arcsec$ or 82 kpc.  Two long filaments reminiscent of tidal tails  extend towards the E and NE. They 
appear as well in a narrow band line-free image obtained with the H$\alpha$/4500+61 filter (not shown here) and emit mostly continuum. These features are evidence for mergers/interactions.

Fig.~21 (top) shows the  H$\beta$-[OIII] 2D spectrum along PA 76.
The [OIII] spatial profile  is dominated by a compact  component with 
FWHM=1.06$\pm$0.02$\arcsec$. The stability of the seeing size during the spectroscopic observations  (FWHM$\sim$ 0.83$\pm$0.03$\arcsec$ 
at the beginning vs. 0.75$\pm$0.01$\arcsec$ at the end)  implies that this central component is spatially
resolved.  In addition,  low surface brightness line emission is detected across a total
extension of of 10$\arcsec$  or 55 kpc. The [OIII] extension towards the West 
overlaps with a region of diffuse, faint continuum emission. Very faint continuum  is also detected  coinciding with the Eastern filament mentioned above.

We have extracted 1D spectra from 4 apertures indicated in Fig. 21 (top). 
The faintest [OIII] extended emission is detected in aperture 4 and
has SB  5$\times$10$^{-18}$ erg s$^{-1}$ cm$^{-2}$ arcsec$^{-2}$. 

The line ratios are plotted in the diagnostic diagram 
in Fig.~21 (bottom).
As expected, the quasar nuclear line ratios are
 better explained by AGN photoionization (the spectrum shows also strong HeII, Table 3). The line ratios of Ap.1 (the adjacent Eastern EELR) 
have large errors because  H$\beta$ is rather noisy, but they are
 very far from the HII galaxies and more consistent with AGN photoionization. Ap. 3 and 4 on the Western EELR  overlap with
the HII galaxy region (only lower limits could be calculated for Ap.4). No useful limits on the line reddening could be estimated. 
     This could be an example of an object where the EELR associated with the quasar is partially ionized by stars, although line reddening should
be estimated before raising conclusions.

This region emits strong continuum, although the origin is uncertain. The [OIII] FWHM values  are 410$\pm$30 km s$^{-1}$, 420$\pm$20 km s$^{-1}$ and 530$\pm$30  km s$^{-1}$ for Ap.1, Ap.2 and Ap. 3 respectively (the  Ap. 4 spectrum is
noisy).  The lines are therefore relatively broad across the whole EELR extension.

Thus, SDSS J1307-02 shows morphological evidence for mergers/interactions. This quasar is associated with an ionized nebula which extends for $\sim$10$\arcsec$ or 55 kpc along PA 76. Although the nuclear gas and the Eastern nebula
are preferentially photoionized by the quasar,  the possibility that the Western nebula  is photoionized by young stars is not discarded.

\begin{figure}
\includegraphics{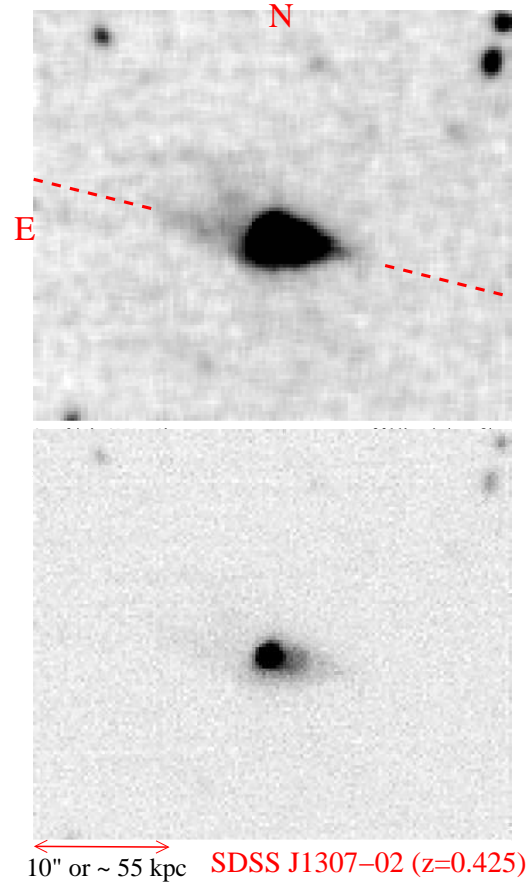}
\vspace{4.5in}
\caption{Broad band  image of the field around SDSS J1307-02. It was obtained with the V\_High  filter and covers  the rest frame spectral range 3462-4327 \AA.   The location of the PA 76 slit is indicated.  The image on the top panel has been smoothed with a 3x3 window. Two filaments stretch towards the E and NE. They are reminiscent of tidal tails. The bottom image (non smoothed) is shown with a different  contrast to highlight  features in the inner, brighter region.}
\end{figure}

\begin{figure}
\includegraphics{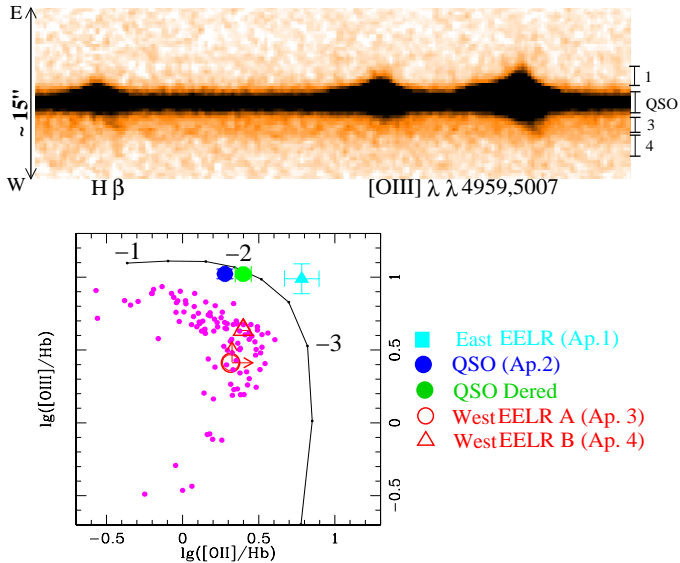}
\vspace{3.2in}
\caption{H$\beta$-[OIII] 2D spectrum of  SDSS J1307-02 along PA 76 (top).The original spectrum has been boxcar smoothed with a 2x2 window. The apertures used for the spectroscopic analysis are indicated on the right side.
The location of the quasar nuclear region (blue and green solid circles)  and the Eastern EELR (cyan solid square) are more consistent with  AGN photoionization 
and lie very far from the HII galaxies. The Western EELR 
(red open circle) overlaps  with the HII galaxy region.  Other symbols and lines as in Fig.~3.}
\end{figure}

\subsection{SDSS J1337-01 ($z=$0.329)}

The [OIII] spatial profile  (Fig.~23) is dominated by an  unresolved component (FWHM=1.40$\pm$0.04$\arcsec$ vs. 1.47$\pm$0.09 seeing). In addition, very faint emission is detected towards the SE along PA 159 extending up to 3.5$\arcsec$ or $\sim$20 kpc from the continuum centroid.
The surface brightness of this EELR is $\sim$10$^{-17}$ erg s$^{-1}$ cm$^{-2}$ arcsec$^{-2}$. The [OIII] spectral profile is unresolved
and has FWHM $\la$170 km s$^{-1}$, compared with the much broader nuclear line (FWHM 930$\pm$20 km s$^{-1}$).
The EELR is shifted by just -25$\pm$20 km s$^{-1}$ relative to the nuclear [OIII] emission.  
The spectrum is too noisy to perform any further kinematic and ionization analysis.  

The nuclear line ratios are shown in Table 3. The H$\beta$ flux is heavily affected by underlying stellar absorption from the galaxy.
The errors on the line ratios and the H$\beta$ flux do not account for this uncertainty. 

The G1 spectrum    shows that it is a continuum source of unknown $z$, maybe a star. 

\begin{figure}
\includegraphics{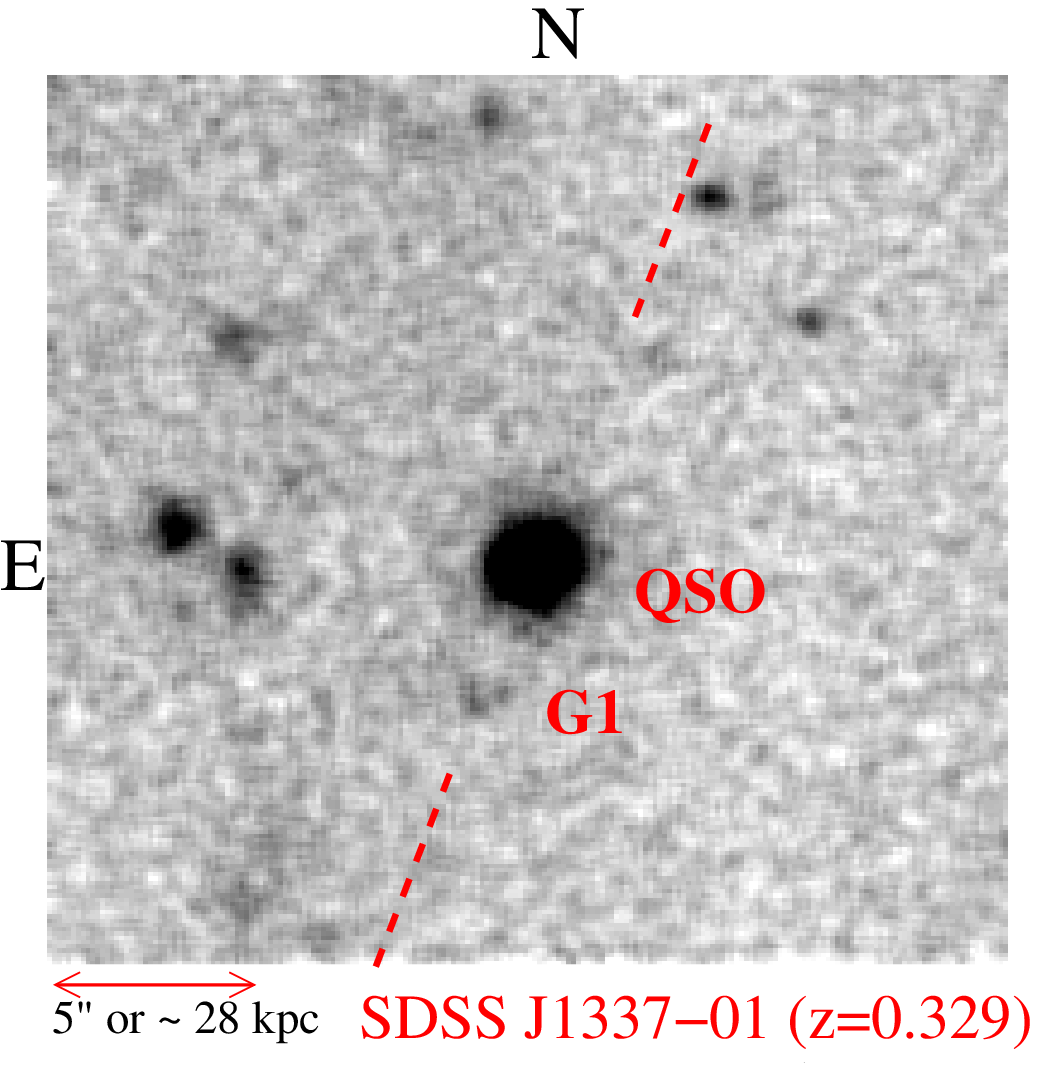}
\vspace{2.5in}
\caption{Narrow band continuum image of SDSS J1337-01 obtained with the  HeII/6500+49 FORS2 filter. It covers the rest frame spectral range
 $\sim$3572-3623 \AA.  The  location of the PA 159  slit is indicated. The original image
has been boxcar smoothed with a 4x4 window. G1 is a continuum source of unknown $z$. }
\includegraphics{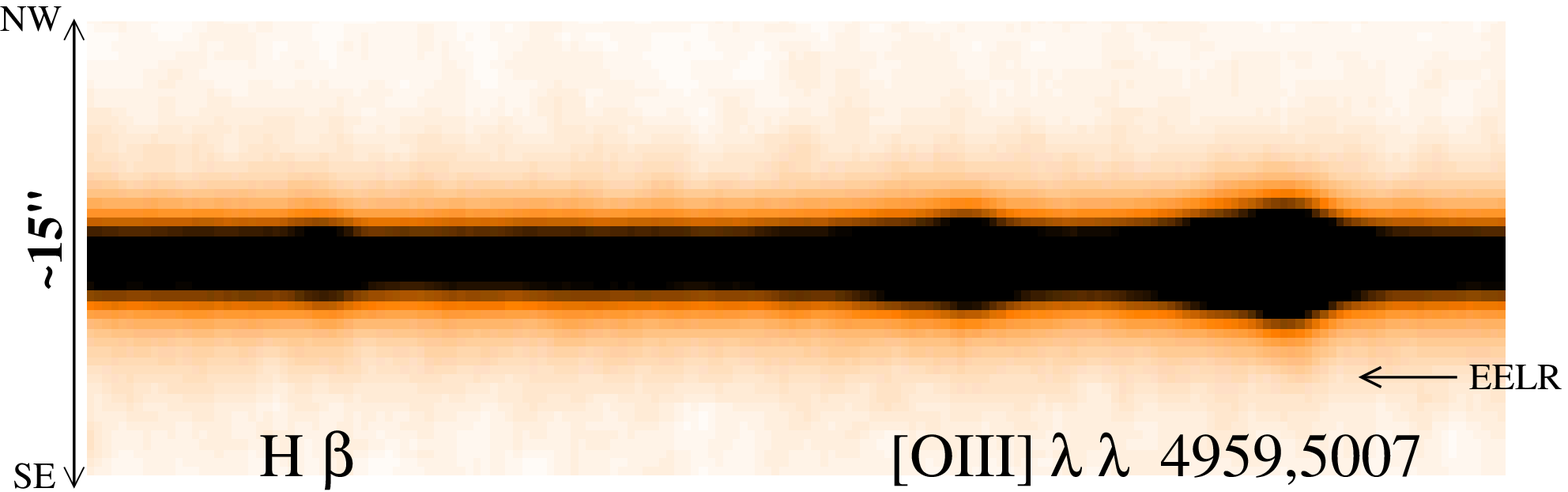}
\vspace{1.5in}
\caption{H$\beta$-[OIII]  2D spectrum of SDSS J1337-01 along PA 159.  The original spectrum
has been boxcar smoothed with a 4x4 window. Very faint extended [OIII] emission is detected up to $\sim$3.5$\arcsec$ from the continuum centroid (indicated with ``EELR'')}
\end{figure}

\subsection{SDSS J1407+02 ($z=$0.309)}

The narrow band continuum image is shown in Fig. 24.
The [OIII]
 spatial profile (Fig.~25)  is dominated by a compact  component of FWHM=0.80$\pm$0.3$\arcsec$ (vs. 0.73$\pm$0.04$\arcsec$), which
is consistent with being unresolved. 
There is no evidence for extended line emission at SB$\ga$3$\sigma$=3.2$\times$10$^{-18}$ erg s$^{-1}$ cm$^{-2}$ arcsec$^{-2}$. 

\begin{figure}
\includegraphics{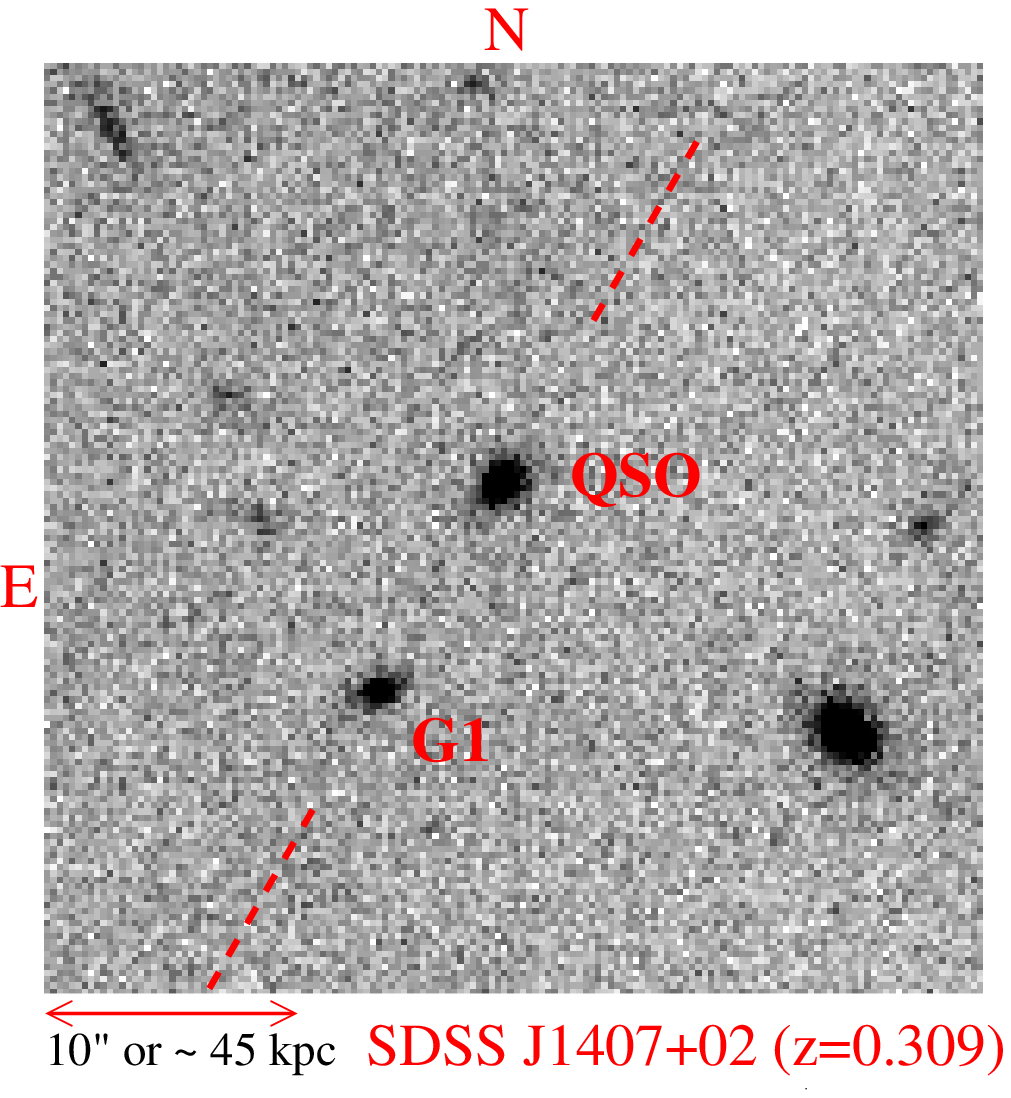}
\vspace{2.6in}
\caption{Narrow band continuum image of SDSS J1407+02   obtained with the HeII/6500+49 FORS2 filter. It covers the rest frame spectral range 3625-3680 \AA.  The PA 150 slit is indicated.  The original image
has been boxcar smoothed with a 2x2 window. G1, which falls also within the slit, is galaxy at $z$=0.254.}
\includegraphics{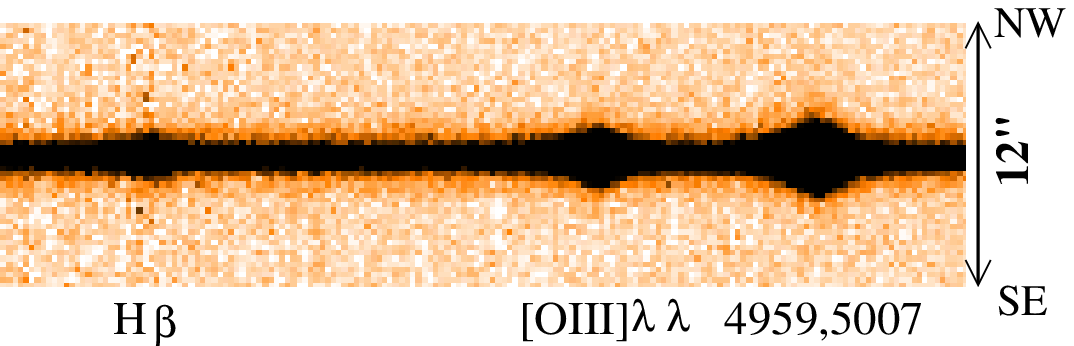}
\vspace{1.2in}
\caption{H$\beta$-[OIII]  2D spectrum of  SDSS J1407+02. There is no evidence for extended line emission along PA 150.}
\end{figure}

The  G1 spectrum  shows that it is a galaxy at $z$=0.254 and therefore unrelated to the quasar.

\subsection{SDSS J1413-01 ($z=$0.380)}

The [OIII] spatial profile  (Fig.~27)  is dominated by
a Gaussian component
of FWHM  1.32$\pm$0.04$\arcsec$, i.e., narrower than the seeing  measured from stars in the images (1.54$\pm$0.09$\arcsec$), indicating that the seeing
improved during
the  spectroscopic exposure.  There is no evidence for extended line emission along PA 45 for SDSS J1413-01 at SB$\ga$3$\sigma$=2$\times$10$^{-18}$ erg s$^{-1}$ cm$^{-2}$ arcsec$^{-2}$.

\begin{figure}
\includegraphics{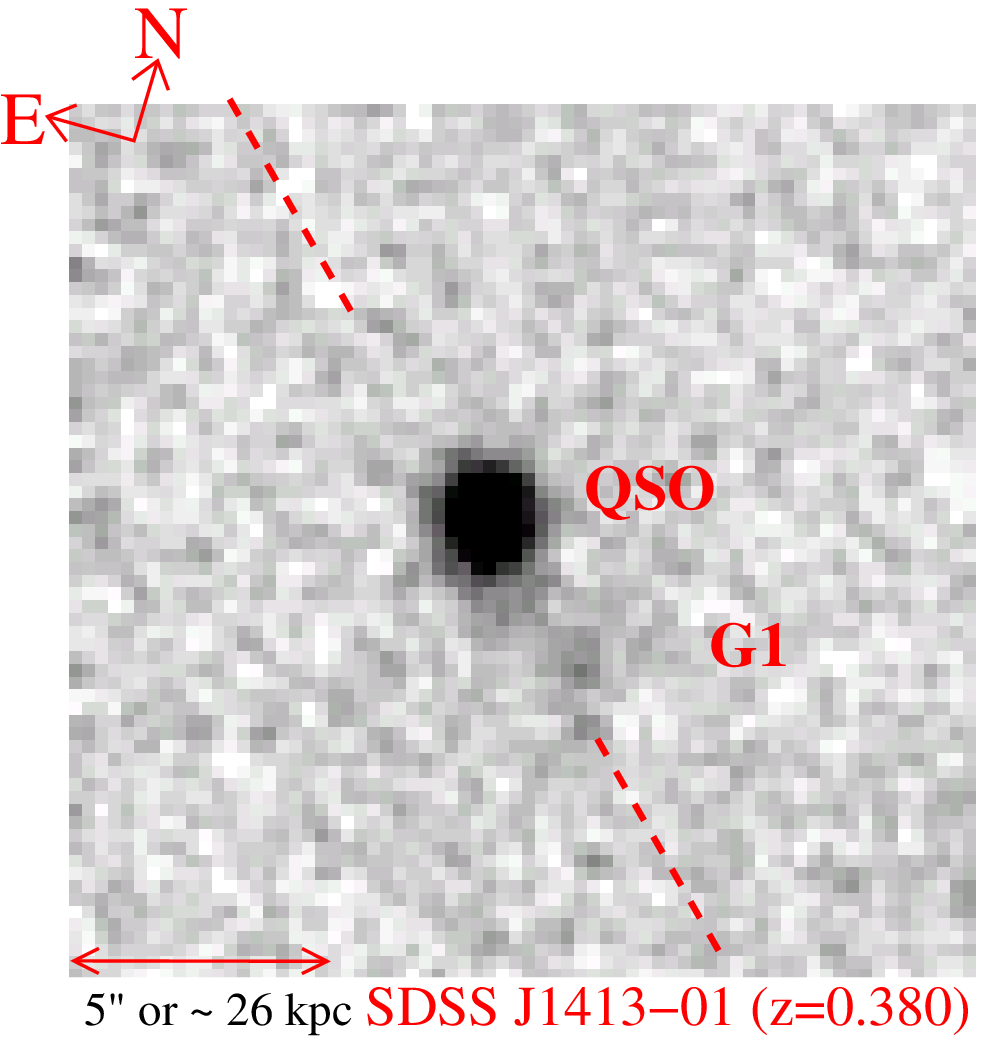}
\vspace{2.4in}
\caption{Broad band  image of  SDSS J1413-01 obtained with the  V\_High  filter
which covers the rest frame spectral range $\sim$3575-4470 \AA. The PA 45  slit  is indicated.  The original image
has been boxcar smoothed with a 2x2 window. G1 is a faint source of unknown $z$. }
\includegraphics{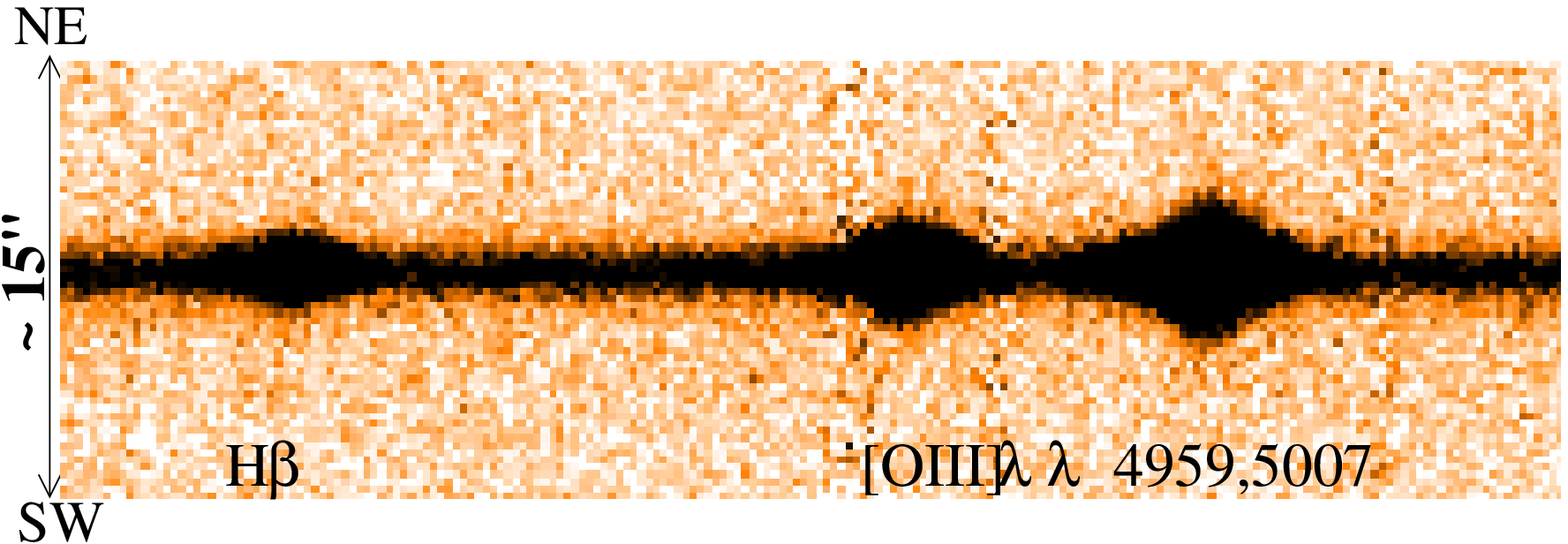}
\vspace{1.3in}
\caption{H$\beta$-[OIII] 2D spectrum of SDSS J1413-01. There is no evidence for extended line emission along PA 45.}
\end{figure}

\subsection{SDSS J1546-00 ($z=$0.383)}

\begin{figure}
\includegraphics{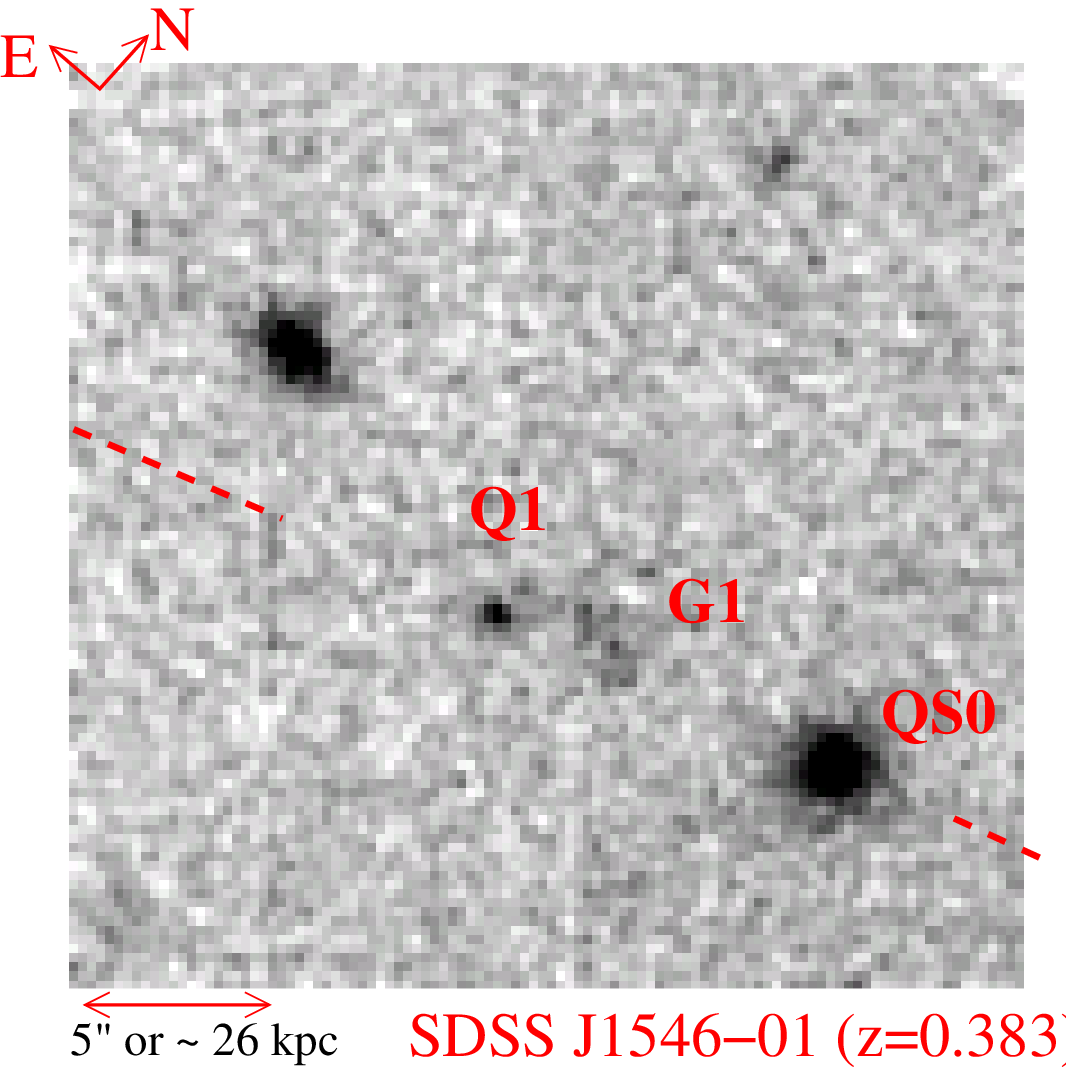}
\vspace{2.6in}
\caption{Broad band V\_High image of the field around SDSS J1546-00 (identified with QSO). It covers the rest frame spectral range $\sim$3570-4460 \AA. The  PA 109 slit location is indicated.  The original image
has been boxcar smoothed with a 2x2 window. G1 and Q1 fall also within the slit.}
\includegraphics{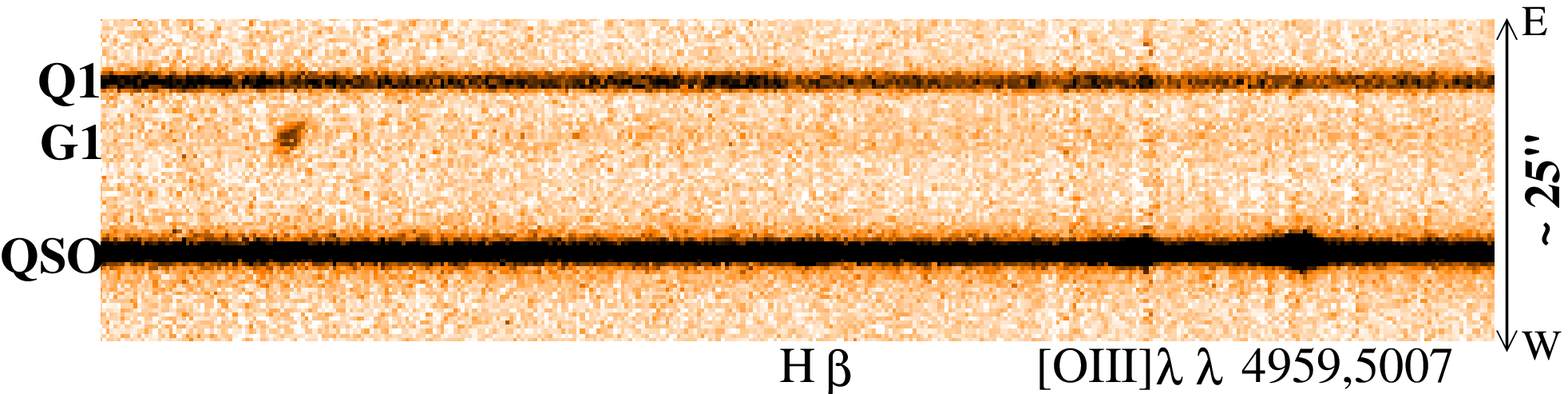}
\vspace{1.5in}
\caption{H$\beta$-[OIII] 2D spectrum of SDSS J1546-00.  Notice the emission line from G1, which
places the object at most probably $z$=0.74. Q1 is a type 1 quasar at $z$=4.12.}
\end{figure}

\begin{figure}
\includegraphics{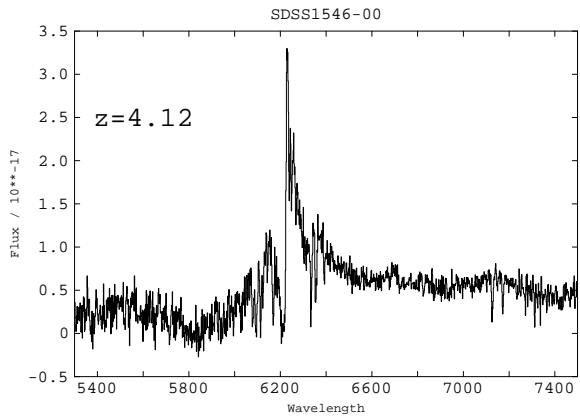}
\vspace{2in}
\caption{Ly$\alpha$ spectrum of the quasar Q1 along the slit of SDSS J1546-00 (Fig.~27 and 28).}
\end{figure}

The [OIII] spatial profile  (Fig.~29) is dominated by
a compact component 
of FWHM=0.97$\pm$0.02$\arcsec$ (vs. seeing FWHM=0.90$\pm$0.01). Taking errors into account and possible
seeing smear during the  spectroscopic exposure,  this is compatible with this component being unresolved. 
There is no evidence for extended line emission along PA 45 for SDSS J1546-00 at SB$\ga$3$\sigma$=3.8$\times$10$^{-18}$ erg s$^{-1}$ cm$^{-2}$ arcsec$^{-2}$.

Q1  (Fig.~28) is a quasar at $z$=4.12 (Fig.30). G1 is an emission line galaxy with a single detected line. It could be [OII]$\lambda$3727 at $z$=0.747 or
  Ly$\alpha$ at $z$=4.36.

The nuclear line ratios are shown in Table 3. The strength of [OIII]$\lambda$4363 relative to [OIII]$\lambda$5007 suggests unusually high electron temperatures (45000$\pm$10000 K).  $T_e>$30000 K is also implied by the SDSS spectrum.

\section{Discussion and conclusions.}

Table 4 is a summary of the results presented in $\delta$3. 
For each object it is specified whether an EELR has been detected and its  nature. When an extended ionized nebula (EIN), other than features such as tidal tails or bridges (see $\delta$3) has been detected, the maximum size measured along the slit  is given in arcsec and kpc. We emphasize that in most cases there is no information available about the possible existence
of extended line emission along other directions.

In  column 5  we specify whether other emission line
features have been found, such as star forming companion galaxies, knots and/or nuclei. In column 6 we say whether there is evidence for galaxy mergers/interactions. The VLT-FORS2  nuclear  [OIII] luminosity and the SDSS [OIII] luminosity  are
 shown in columns 8 and 9. We have added the information on SDSS J0123+00 ($z=$0.399), observed as part of the same observational programme (Villar-Mart\'\i n et al. \citeyear{vil10}). 

\begin{table*}
\centering
\begin{tabular}{llllllllll}
\hline
Object  &  EELR & Nature &  Max. Size & Max. Size  & Other features? & Interactions? & SF? &  L([OIII])$_{nuc}$ & L([OIII])$_{SDSS}$ \\ 
        &  &  & of IN (")  & of IN (kpc) & & &  \\ 
(1) & (2) & (3) & (4) & (5) & (6) & (7) & (8) & (9)  & (10) \\ \hline
SDSS J2358-00 & Yes & EIN & 12 & 64 & SFCG & Yes & Yes & 9.27 & 9.32 \\
SDSS J0025-10 & Yes  &  TT  & --  &  -- &   SFCN,SFCK  & Yes &  Yes &  8.60  & 8.73 \\
SDSS J0217-00 &  Yes &  EIN & 4 & 19 & SFCN & Yes & Yes & 8.81 & 8.75 \\
SDSS J0234-00  & No &  -- & --   &  -- & No & No  & No & 8.73 & 8.77  \\ \hline \hline
SDSS J0849+01 &  No & -- & -- & --  & No & No &  No & 8.14  & 8.06  \\ 
SDSS J0955+03 & Yes & EIN & 7.5 &  38.5     & No  & No  & No & 8.04 & 8.60 \\ 
SDSS J1153+03 & Yes &  EIN & 8 & 52 & SFCG & Yes & Yes &  9.54 & 9.61 \\ 
SDSS J1228+00 &   No & -- &  --   &  --   & No & No  & No  & 9.06 & 9.28 \\ 
SDSS J1307-02 & Yes & EIN &  10 & 55 & No & Yes &  ? & 8.91 & 8.92  \\
SDSS J1337-01 &  Yes & EIN &   3.5 & 20 & No & No  & No &  8.50 & 8.72  \\ 
SDSS J1407+02  & No & --   -- & & --    & No & No & No & 9.23 & 8.90 \\ 
SDSS J1413-01  & No & --   & --  & --   & No & No &  No & 8.97 & 9.25 \\ 
SDSS J1546-00  & No  &  --   & -- &  -- & No & No  & No &  8.24 & 8.18 \\ \hline
SDSS J0123+00* & Yes & TB,EIN & 34 & 180 & SFCG? & Yes & Yes & 9.02 & 9.13 \\
 \hline
\end{tabular}
\caption{Summary of results. For each object it is specified whether an EELR has been detected or not (column 2) and its nature: EIN (extended ionized nebula), TT (tidal tail), TB (tidal bridge).
 When an IN is detected, the maximum measured size is given in  $\arcsec$ and kpc (columns 4 and 5).  Column 6 indicates whether other emission line features have been found
 and their nature: SFCG (star forming companion galaxy), SFCN (star forming companion nucleus), SFCK (star forming knot). In Columns 7 and 8 it is said whether the object shows signs of mergers/interactions (7) and  star formation (8). The nuclear [OIII] luminosity measured from the VLT-FORS2 spectra
is given in column 9 relative to the solar luminosity and in log. The [OIII] luminosity measured from the SDSS spectra (3$\arcsec$ fibers) is given in column 10
(taken from Zakamska et al. 2003).  
 See Villar-Mart\'\i n et al. (2010) for results on SDSS J0123+00. The objects observed in the 2008 and 2009 runs are separated by a double horizontal line.}
\end{table*}

The  summary of our results is presented next. As we mentioned in $\delta$3, they cannot be generalized to all
optically selected type 2 quasars, due to different possible biases.

\begin{itemize}

\item Evidence for galaxy mergers/interactions is found in 6/14 objects  (SDSS J2358-00, SDSS J0217-00, SDSS J0025-10, SDSS 1153+03, SDSS J1307-02, SDSS J0123+00).
 Such evidence was suggested by other authors  for SDSS J2358-00 and  SDSS J0123+00 (Zakamska et al. \citeyear{zak06}) based on HST images.

The detection rate in our sample is possibly higher.   Our images are rather shallow for  most  objects and  the rate of detection 
of the characteristic morphological features indicative of mergers/interactions depends on the depth of 
the images.
For comparison, this rate  is very high in powerful radio galaxies (75\%-95\%, Ramos Almeida et al. 2010).

\item Evidence for recent star formation  in the neighborhood of the quasars is found for 
5/14 objects. The star formation is happening in general in companion galaxies 
(SDSS J2358-00, SDSS 1153+03, SDSS J0123+00?),  knots (SDSS J0217-00, SDSS J0025-10) and/or
a  nuclei (SDSS J0025-10). Star formation is possibly happening as well in the tidal bridge that connects SDSS J0123+00 with its interacting companion.
 All these objects show evidence for mergers/interactions, which could be responsible for  triggering the star formation activity.

 As mentioned earlier, AGN photoionization cannot be completely discarded in favour of stellar photoionization based only on the line ratios studied here. Other properties, such as the narrowness of the emission lines and the spatial morphology  favour the stellar photoionization scenario. This is further supported by the fact that those  regions with line ratios consistent with HII galaxies are in general companion galaxies, nuclei,  knots and tidal features  associated with quasars which are undergoing a merger event, where active star formation is naturally expected. In addition, if the gas was photoionized by the AGN, this would mean that in all cases the companion knots/nucei/galaxy/tidal tails are within the quasar ionization cones, which is quite unlikely.

SDSS 1307-02 also shows morphological evidence for mergers/interactions. This might be
another object  with star formation, although it is  necessary to determine the impact of line reddening
before raising a definitive conclusion.

Therefore, we find a diverse population of systems. Several (particularly those with high surface
brightness off-nuclear structures indicative of mergers) are composite
in their emission line properties showing a combination of AGN and
star formation features. It is possible that at least some of
the radio-quiet type 2 quasar  are triggered in galaxy major
interactions involving at least one gas-rich galaxy, although it is essential to compare the rate of detection of mergers/interactions with that of a non active galaxy control sample.

Since at least some 
of the star forming  regions of the type 2 quasars would be included in the 
SDSS aperture, this might explain the type 2 quasars with composite AGN/star formation spectra 
(Villar-Mart\'\i n et al. \citeyear{vil08}). 

This result seems different from that obtained for powerful radio
galaxies in which, based on the  analysis of the extended line emission, there seems to 
be less such evidence. Although ~20-30\% of radio galaxies show 
evidence for star  formation from their continuum properties, in many of 
these cases the young stellar populations are relatively old and do not produce significant 
emission lines (Tadhunter et al. \citeyear{tadh05},\citeyear{tadh11}). The radio galaxy PKS 1932-464 at $z$=0.23 (Villar-Mart\'\i n etal. \citeyear{vil05}) is a
remarkable exception.

We do not know whether star formation is ongoing or has recently happened {\it within} the host galaxies of the quasars studied here, except for SDSS J1337-01, where 
Bian et al.  (\citeyear{bian07}) found a very young stellar population.  Intense star formation activity would not be surprising, since different works suggest that radio quiet type 2 quasars at similar $z$ have very high star-forming luminosities, much higher than  field galaxies
(e.g. Lacy et al. \citeyear{lacy07}, Zakamska et al. \citeyear{zak08},  Hiner et al. \citeyear{hin09}).

\item 8/14 objects  have EELRs. In most cases (7/8) the EELR apparently  consists of an extended ionized nebula associated with the quasar. For one object (SDSS J0123+00), 
moreover, a tidal bridge is also part of the EELR.  For another quasar, the EELR consists of a tidal tail (SDSS J0025-10). The detection rate of EELRs in our sample
is a lower limit,
since the results presented here refer in most cases to a single  blind spatial direction.

The sizes of the extended ionized nebulae vary between several kpc and up to 64 kpc (180 kpc, including SDSS 0123+00).
At the detection limit of our data, the non detection of EELRs implies that, if existing, the size along the studied direction 
is $\leq$few kpc. \cite{green11}  find  that ionized gas is ubiquitous
within the quasar host galaxies but rare at larger spatial scales ($\ga$10 kpc). However, these authors reach much shallower surface brightness levels $\sim$10$^{-16}$ erg s$^{-1}$ cm$^{-2}$ arcsec$^{-2}$.

Humphrey et al. (2010) found EELRs associated with 2 out of 6 type 2 SDSS quasars with maximum total sizes of $\sim$40 and 27 kpc respectively . We emphasize that these data were $\ga$10 times shallower that those presented here.  

As we explained in $\delta$1, studies of type 1 quasars show that luminous (L[OIII]$>$5$\times$10$^{41}$ erg s$^{-1}$) EELR are preferentially associated with steep spectrum radio loud quasars with high nuclear line luminosities (L[OIII]$_{nuc}>$6.5$\times$10$^{42}$ erg s$^{-1}$, 
Fu \& Stockton \citeyear{fu09}, \citeyear{fu07}). 
Our data are too limited to make a proper comparison, since we would  require a complete spatial coverage around the type 2 quasars and an accurate measurement of the total EELR [OIII] luminosity. However, there are two objects which are worth mentioning: 
SDSS J0123+00 and SDSS J2358-00. The [OIII] EELR luminosities measured from the spectra (therefore, these are lower limits) are (3.05$\pm$0.08)$\times$10$^{42}$ erg s$^{-1}$ and
(5.1$\pm$0.1)$\times$10$^{41}$ erg s$^{-1}$ respectively. Neither object is radio loud; therefore, these two examples demonstrate
that a luminous EELR is not necessarily associated with a powerful radio source.

The nuclear [OIII] luminosities are 7.2$\times$10$^{42}$ erg s$^{-1}$ and 4.4 $\times$10$^{42}$ erg s$^{-1}$ for SDSS2358-00 and SDSS J0123+00
respectively 
 (we have tried to use an aperture
similar in physical size to that used by Fu \& Stockton \citeyear{fu09}). Both are close to the minimum nuclear L[OIII]  found by these authors for quasars with luminous EELRs\footnote{The comparison of the [OIII] nuclear luminosities should be considered with caution, since [OIII] has high critical densities and it can be partially produced in  regions which are not observable for type 2 quasars. I.e. the [OIII] line luminosity
 depends on orientation (e.g. di Serego Alighieri et al.  \citeyear{spe97}).  According to this, for a type 1 orientation (as Fu \& Stockton's objects)
we would measure a higher nuclear [OIII] luminosity.}.

\item We have  investigated the gas ionization mechanism in the EELRs  of 5 quasars where this study was possible. Stellar photoionization is 
very probably present in the tidal tail and tidal bridge of SDSS J0025-10 (this work) and SDSS J0123+00 (Villar-Mart\'\i n et al. \citeyear{vil10}) respectively; maybe also in the EELR of SDSS J1307-02. The extended ionized nebulae are preferentially photoionized by the quasar  in at least 4  of them  (SDSS J2358-00, SDSS 1153+03, SDSS J1307-02, SDSS J0123+00). 

\item The  nuclear gas is preferentially photoionized by the active nucleus.   The H$\beta$ luminosities measured for  the quasars in our sample are in the range $L_{H\beta}\sim$(0.2-9.3)$\times$10$^{41}$ erg s$^{-1}$. Similar values have been measured for the NLR of some  radio quiet type 1 quasars (e.g. Bennert et al. \citeyear{benn02}).
The implied photon ionizing lumonisity $Q_{ion}$ {\it absorbed} by the gas is given by $Q_{ion}=\frac{L_{H\beta}}{h~\nu_{H\beta}} \times \frac{\alpha_B}{\alpha_{H\beta}^{eff}}$ 
where $h\nu_{H\beta}$ is the energy of a H$\beta$ photon and $\alpha_B$ and $\alpha_{H\beta}^{eff}$ are the total and effective H$\beta$ case B Hydrogen recombination coefficients (Osterbrock \citeyear{ost89}). 
$Q_{ion}$ is in the range $\sim$(0.08-4)$\times$10$^{54}$ s$^{-1}$.

In the previous section, we derived  log($U$) for 4 objects from the AGN models, which is in the range -1.3 to -2.0.   Using the corresponding L(H$\beta$) values  and $U=\frac{Q}{4~\pi~r^2~n~c}$ ($\delta$3), a  $n=$100 cm$^{-3}$ density implies $r\sim$500 pc-1.5 kpc. NLR sizes of type 1 quasars measured from HST images are $\sim$1-several kpc  (Bennert et al. \citeyear{benn02}).  It should be kept in mind that the $r$ values we have inferred refer to the region that emits the bulk of the line emission, while the actual radius of the NLR could be larger. 

\item Forbidden high ionization [FeVII]  lines have been detected in the nuclear spectrum of one object (SDSS J1153+03). Very broad underlying H$\beta$ and strong continuum are also detected. Based on previous works
by other authors, we propose that this object is seen with a viewing angle intermediate between pure type 1 and type 2 orientations so that the illuminated face of the torus and regions interior to it can be at least partially observed. Emission from the intermediate region between the NLR and the BLR  is possibly detected from SDSS J1228+00 as well.
These are the highest $z$ objects in our sample ($z=$0.575).

\end{itemize}

\section*{Acknowledgments}
This work has been funded with support from the Spanish Ministerio de Ciencia e Innovaci\'on through the grants AYA2004-02703,
AYA2007-64712 and AYA2009-13036-C02-01 and co-financed with FEDER funds.  We thank the referee for the thorough revision of the original manuscript 
and very useful comments that helped to improve the paper. Thanks to the staff at Paranal Observatory for their support during the observations.

\end{document}